\documentclass[superscriptaddress,12pt,tightenlines,aps,nofootinbib,pre]{revtex4-1}
\usepackage{color,epsfig,graphics,amssymb,amsmath,subeqnarray,graphicx,amsthm,subfigure,mathrsfs,bm,color,xcolor,soul}

\def\bea{\begin{equation}}
\def\eea{\end{equation}}

\newcommand{\diff}[2]{\frac{\mbox{d} #1}{\mbox{d} #2}}

\newcommand{\pdiff}[2]{\frac{\partial #1}{\partial #2}}

\newcommand{\soltype}[1]{\textsc{\romannumeral #1}}

\newcommand{\JackRed}[1]{{#1}}

\begin{document}

\title{The influence of invariant solutions on the transient behaviour of an air bubble in a Hele-Shaw channel.}

\author{J. S. Keeler}
\email{jack.keeler@manchester.ac.uk}
\affiliation{School of Mathematics and Manchester Centre for
   Nonlinear Dynamics (MCND), University of Manchester, Oxford Road, Manchester, M13 9PL, UK}

\author{A. B. Thompson}
\email{alice.thompson@manchester.ac.uk}
\affiliation{School of Mathematics and Manchester Centre for
  Nonlinear Dynamics (MCND), University of Manchester, Oxford Road, Manchester, M13 9PL, UK}

\author{G. Lemoult}
\email{gregoire.lemoult@univ-lehavre.fr}
 \affiliation{Normandie Universit\'{e}, UniHavre, CNRS, UMR 6294, Laboratoire Onde et Milieux Complexes (LOMC) 53, rue de Prony, 76058 Le Havre Cedex, France.}

\author{A. Juel}
\email{anne.juel@manchester.ac.uk}
\affiliation{School of Mathematics and Manchester Centre for
   Nonlinear Dynamics (MCND), University of Manchester, Oxford Road, Manchester, M13 9PL, UK. School of Physics and Astronomy and MCND, University of
   Manchester, Oxford Road, Manchester, M13 9PL, UK}

\author{A. L. Hazel}
\email{andrew.hazel@manchester.ac.uk}
\affiliation{School of Mathematics and Manchester Centre for
  Nonlinear Dynamics (MCND), University of Manchester, Oxford Road, Manchester, M13 9PL, UK}




\email{jack.keeler@manchester.ac.uk}

\vspace{1cm}

\begin{abstract}
We hypothesize that dynamical systems concepts
used to study the transition to turbulence
in shear flows are applicable to other transition
phenomena in fluid mechanics. In this paper, we
consider a finite air bubble that propagates within a
Hele-Shaw channel containing a depth-perturbation. Recent experiments revealed that
the bubble shape becomes more complex, quantified
by an increasing number of transient bubble tips, with
increasing flow rate. Eventually the bubble changes
topology, breaking into two or more distinct entities
with non-trivial dynamics. We demonstrate that
qualitatively similar behaviour to the experiments is
exhibited by a previously established, depth-averaged
mathematical model; a consequence of the model's
intricate solution structure. For the bubble volumes
studied, a stable asymmetric bubble exists for all flow
rates of interest, whilst a second stable solution branch
develops above a critical flow rate and transitions
between symmetric and asymmetric shapes.
The region of bistability is bounded by two Hopf bifurcations on the second
branch. By developing a method for a numerical weakly
nonlinear stability analysis we show that unstable
periodic orbits emanate from the Hopf bifurcation at
the lower flow rate and, moreover, that these orbits
are edge states that influence the transient behaviour of
the system.
\end{abstract}
%

\maketitle
  
\section{Introduction}\label{sec:intro}

A Hele-Shaw channel consists of two parallel glass plates,
separated by a distance much smaller than the width of the
channel. If a trapped viscous fluid is extracted at a constant flux
from one end of the channel and an air bubble is placed at the other
end, then the bubble will propagate and change shape as it does so.
For sufficiently large bubbles the only stable
solution is for the bubble to propagate symmetrically along the
centreline of the channel; a solution analogous to the
symmetric semi-infinite air finger that develops when one end
of the channel is left open to the atmosphere
\cite{saffman1958penetration}. For higher flow rates in large aspect
ratio channels, propagating air fingers
develop complex patterns via multiple tip splitting
events as well as side-branching
\cite{Tabeling1987,MooreJuel2002}. The onset of this complex interfacial dynamics
appears to be a subcritical transition; a feature that it
shares with the transition to turbulence in shear
flows. Specifically, the steady symmetrically propagating solution is linearly
stable for all values of the flow rate at which it has been
computed, meaning that finite perturbations are required to initiate
the complex dynamics. Moreover, the value of the critical dimensionless flow rate for the onset of 
patterns cannot be precisely determined and is very sensitive to the level of perturbation in the
system: the transition occurs at
lower dimensionless flow rates as the roughness of the channel walls
is increased\cite{Tabeling1987}.

\begin{figure}
  \centering
  \includegraphics[scale=0.8]{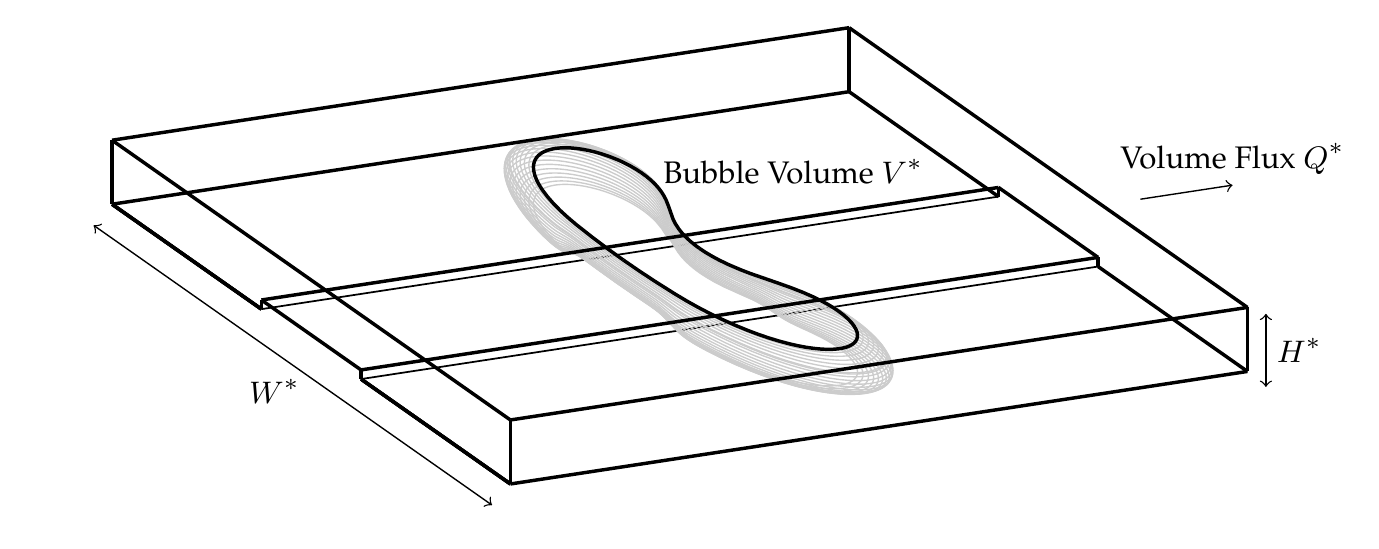}
    \caption{Sketch of a bubble propagating in a Hele-Shaw  channel
      with a depth-perturbation on the bottom, showing the 3D location
      of the interface, depth-perturbation and direction of
      propagation and the dimensional quantities, $W^*$, $H^*$, $V^*$
      and $Q^*$.}
    \label{fig:dimdomain}
    
\end{figure}

\begin{figure}
  \centering \includegraphics[scale=0.85]{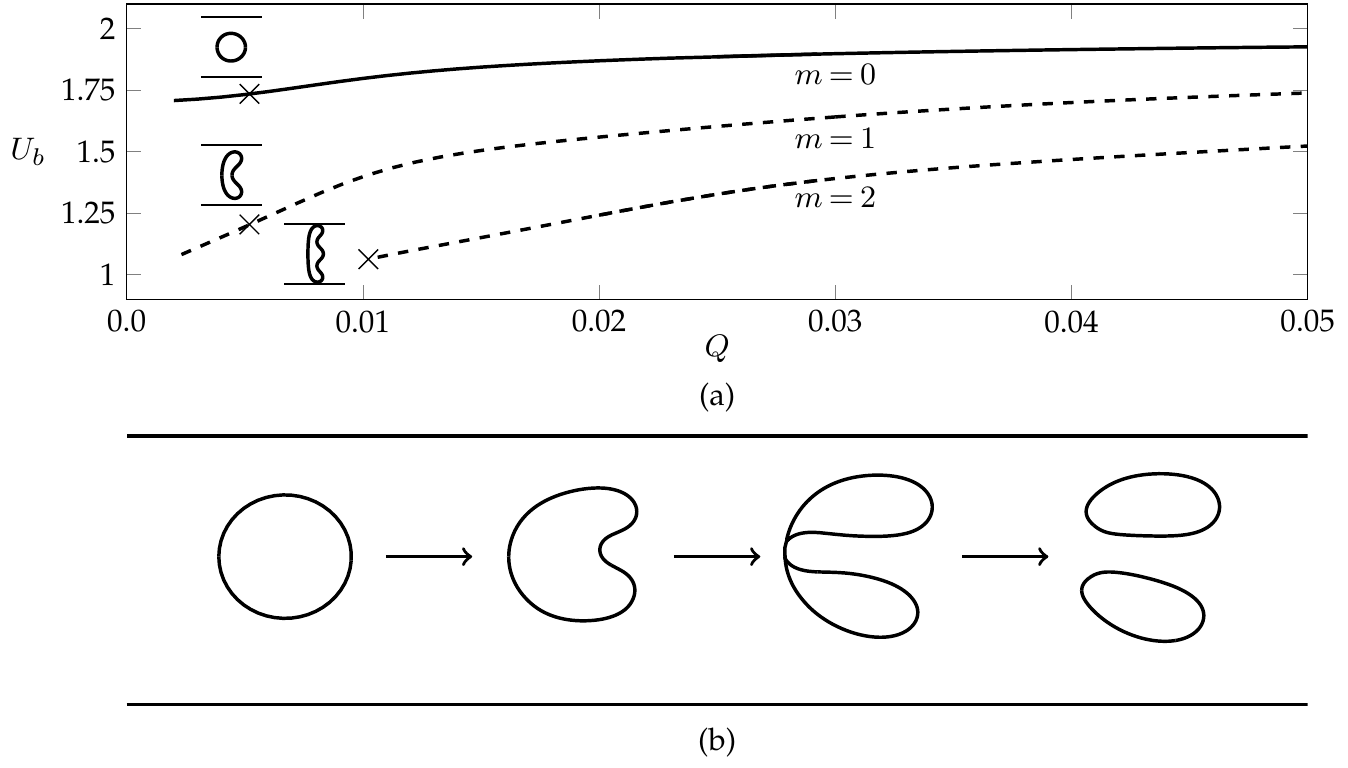}
  \caption{(a) The steady solution space for a bubble of fixed volume in an unoccluded channel with side walls. The solution
    measure $U_b$ is the speed of bubble relative to fluid ahead,
    while $Q$ is the dimensionless imposed flow rate. The insets
    indicate the bubble shapes denoted by crosses on the solution
    branches and are on a 1:1 scale with $x,y\in[-1,1]$. These solutions
    were calculated using the set of equations described in
    \S~\ref{sec:form}. (b) A sketch of an initial bubble evolving in
    time to form two tips and eventually break up, yielding two
    separate bubbles.}
  \label{fig:break_up1}
  
\end{figure}

In this paper, we concentrate on bubble propagation in a
geometrically-perturbed Hele-Shaw cell: a rectangular prism
is added to the base of the channel, as sketched in figure~\ref{fig:dimdomain}.
By studying bubbles, rather than air fingers, and working in a co-moving frame, we can follow long-time evolution of the
system in short computational domains. This relatively simple system still exhibits a wide variety of nonlinear dynamical phenomena
\cite{de2009tube,pailha2012oscillatory,hazel2013multiple,jisiou2014geometry,thompson2014multiple,franco2016sensitivity,franco2017propagation,
    franco2017bubble} and the recent experimental results of Franco-G{\'o}mez \textit{et
    al}\cite{franco2017propagation} have shown that the system contains regions of
  bistability in which finite perturbations can provoke a
  range of possibilities for the time-evolution of an initially centred
  bubble. For small volume
  fluxes the bubble will eventually settle towards a stable asymmetric
  state on one side (or the other) of the depth perturbation,
  but if the volume flux exceeds a critical threshold the bubble
  shape becomes increasingly deformed before eventually breaking up
  into two or more distinct parts, as sketched in figure
  \ref{fig:break_up1}(b), despite theoretical predictions that a stable
  steady state exists for
  all flow rates; a feature preserved from the unperturbed Hele-Shaw
  channel. Thus, this model system allows us to explore whether the
  dynamical systems concepts applied in the study of
  transition to turbulence in shear flows also apply to the transient
  behaviour of other canonical
  problems in fluid mechanics.

  We study the behaviour theoretically by finding invariant
  solutions, namely steady states and periodic orbits, of the system.
  We pursue the idea, hypothesised
  by Franco-G{\'o}mez \textit{et al} \cite{franco2017propagation}, that
  when the flux is large enough the complex time-dependent behaviour of
  a single bubble can be interpreted as a transient exploration of the
  stable manifolds of weakly unstable \textit{edge states}. We consider
  only the dynamics before the changes in topology when the bubble breaks up into two or more
  separate bubbles: systems with their own dynamics that we do not pursue here.

  In the context of fluid mechanics an edge state is
an invariant solution of the governing equations that has only a
small number of unstable eigenvalues and whose stable manifold forms the `boundary' between
two qualitatively different dynamical outcomes. Edge tracking techniques have been
utilised in a number of different scenarios including shear flow and
pipe flow \cite{kerswell2005recent,schneider2007turbulence,schneider2008laminar,eckhardt2008dynamical}
and more recently droplet breakup \cite{gallino2018edge}. In these
studies the edge state was found by direct numerical simulation of the
governing equations and interval bisection of the initial
conditions. This methodology is easy to implement numerically but can
be computationally expensive and does not reveal whether
the edge state corresponds to a steady state, periodic orbit or
other invariant solution.

   One advantage of the Hele-Shaw system is that the
only nonlinearities in the system arise due to boundary conditions
on the bubble because the reduced Reynolds number is extremely small.
Thus the shape of the interface gives an immediate visual
representation of nonlinear behaviour. Another advantage is that
by assuming that the width of the channel is large compared to its
height, the behaviour of the system can be described by
a depth-averaged set of equations that is more
amenable to analysis than the full Navier--Stokes
equations\cite{saffman1958penetration,taylor1959note}.
These reduced equations (stated in \S~\ref{sec:form}), often called Darcy or Hele-Shaw
equations, have been used in most theoretical studies of this system, and
lead to predictions for the bubble shape as viewed from above
(sketched in figure~\ref{fig:break_up1}(b)). In the unperturbed
system at zero surface tension,
exact steady solutions can be found by conformal mapping techniques,
and for each bubble volume, there is a two-parameter family of
solutions described by the centroid offset and bubble speed
\cite{tanveer1987surfacetension,tanveer1987stability,tanveer1987newsolutions}. The
introduction of surface tension selects both a main, symmetric branch
of linearly stable steady solutions which persists for all fluxes,
along with a countably infinite sequence of unstable `exotic' bubble
shapes; these exist in both channels bounded by parallel side walls
\cite{tanveer1987stability} and in unbounded channels
\cite{green2017effect}. Figure~\ref{fig:break_up1}(a) shows the bubble
shape and speed for the first three solution branches in channels with
a rectangular cross-section. Introducing a
depth-perturbation to the system (see sketch in
figure~\ref{fig:dimdomain}) allows the solution branches to interact,
which results in the diverse range of observed stable steady states and
time-dependent behaviour
\cite{de2009tube,pailha2012oscillatory,hazel2013multiple,jisiou2014geometry,thompson2014multiple,franco2016sensitivity,franco2017propagation,franco2017bubble}. 

  We will show that the bifurcation structure of the depth-averaged equations does indeed feature
bistability between steady states, and that previously unknown unstable periodic orbits influence
whether the bubble breaks up or returns to a stable
configuration. We have sufficient knowledge of the solution structure that, rather than edge tracking, we can
develop a general numerical procedure for weakly nonlinear analysis that allows us to determine
an approximation to the periodic orbits and steady states. We
emphasise that the method described in this paper, although applied to
our specific set of equations, can readily be adapted for other
problems and we provide a numerical recipe for this procedure.
The results of the weakly nonlinear analysis, along with time-dependent calculations, 
show that the previously unknown unstable periodic orbits are indeed edge states of the system that
influence the eventual fate of the bubble.


The paper is set out as follows. Initially, in \S~\ref{sec:form}, we
introduce the details of the system and describe the depth-averaged
governing equations, non-dimensional parameters and numerical
methods. In \S~\ref{sec:one}, we start our investigation with a range
of initial-value calculations and compare these to the previous
experiments \cite{franco2017propagation}.
In \S~\ref{sec:two}, we present an extended bifurcation diagram for
the invariant steady states and perform a linear stability
analysis. In \S~\ref{sec:four} we derive our general method for
determining the weakly nonlinear approximation to the invariant
periodic orbits of the system by employing the method of multiple
scales near the bifurcation points. Finally, in \S~\ref{sec:six}, we
summarise our results and briefly discuss the importance of the
periodic orbits to the transient behaviour of the system and their
interpretation as edge states.

\section{Governing Equations}\label{sec:form}
\begin{figure}
  \centering
  \includegraphics[scale=0.85]{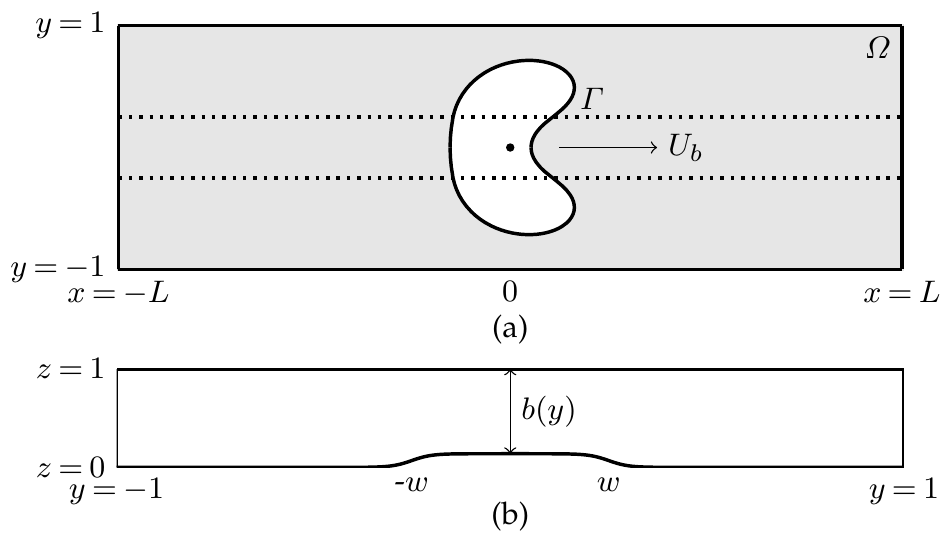}

    \caption{The non-dimensional computational domain. (a) Top-view
      sketch of bubble propagation, which is the focus of the
      depth-averaged model.  Calculations take place in a frame of
      reference moving with the bubble centroid, and the computational
      domain is truncated in $x\in[-L,L]$ where $L$ at a dimensionless
      distance, typically chosen to be $L=4$ in the simulations.
    (b) For the depth-averaged calculations, a smoothed version of the
      depth-perturbation is used, plotted here in the $(y,z)$ plane
      when $h=0.024,s=40,w=0.25$.
    \label{fig:nondimdomain}
    }
\end{figure}


The physical situation is shown in figure~\ref{fig:dimdomain}. The
channel has outer dimensional width $W^*$ and height $H^*$.
The channel is filled with a viscous fluid of density $\rho$ and
dynamic viscosity $\mu$, containing a single air bubble with known
volume $V^*=(W^*)^2H^*V/4$, where $V$ is the non-dimensional volume. Fluid flow is driven by withdrawing fluid at a constant
flux $Q^*$ far ahead of the bubble. The system is non-dimensionalised
based on a length scale $W^*/2$ in the $x$ and $y$ directions, and on
the velocity scale $U_0^* = Q^*/(W^*H^*)$. We define the
non-dimensional flow rate
%
%
%
$ Q = \mu U_0^*/\gamma, $ where $\gamma$ is the coefficient of surface
tension at the air/fluid interface. Additionally we define the aspect
ratio of the channel $\alpha=W^*/H^*$. Details of the derivation of
the depth-averaged equations from the Navier-Stokes equations can be
found in \cite{thompson2014multiple,franco2017propagation}.  The
critical assumptions are that the channel aspect ratio $\alpha$ is
large; the reduced Reynolds number $U_0^* W^*/(\rho \mu\alpha^2)$ and
Bond number $\mbox{Bo} = \rho gL^{*2}/\gamma$ are small; the
bubble occupies the full height of the channel; and the component
of curvature as $z$ varies corresponds to a semi-circle filling the
channel height {with the fluid perfectly wetting} the upper
and lower walls.  We neglect the thin film corrections proposed by
Homsy \cite{homsy1987viscous} and Reinelt \cite{reinelt1987interface},
because 
%
they do not change the qualitative comparison with the experiment, and
it is not obvious how they should be modified due to the presence of
the depth-perturbation and multiple solutions
\cite{thompson2014multiple,franco2016sensitivity,franco2017propagation}.

The nondimensional domain is shown in
figure~\ref{fig:nondimdomain}. The depth-averaged velocity at any
point within the fluid is given by $\mathbf{\hat{u}} = -b^2(y) \nabla
p$, where $p=p(x,y)$ is the fluid pressure and $b(y)$ is the height of
the channel (see figure~\ref{fig:nondimdomain}(b)). We work in a frame
moving at dimensionless speed $U_b(t)$ chosen at each time step so
that the $x$ component of the bubble centroid remains fixed in this
moving frame. The equations to be solved are
\begin{subequations}
  \begin{align}
    \nabla \cdot(b^3(y)\nabla p) = 0\qquad \mbox{in\ }\:&
    \Omega,\label{exact_eqn1}\\ p_b - p = \frac{1}{3\alpha
      Q}\left(\frac{1}{b(y)}+\frac{\kappa}{\alpha}\right)\qquad
    \mbox{on\ }\:&
    \Gamma,\label{exact_eqn2}\\ \textbf{n}\cdot\textbf{R}_t +
    \textbf{n}\cdot \mathbf{e}_x U_b(t) + b^2(y)\textbf{n}\cdot\nabla
    p =0\qquad \mbox{on\ }\:&
    \Gamma,\label{exact_eqn3}\\ \frac{\partial p}{\partial y} =0
    \qquad \mbox{on\ } \:& y =\pm 1, \label{exact_eqn4}\\
     p=0 \qquad \mbox{on\ } \:& x =
     -L, \label{inlet_equation}\\ \frac{\partial p}{\partial x} = -1
     \qquad \mbox{on\ } \:& x = L, \label{exact_eqn5}\\ \int b(y)
     \textbf{R}\cdot\textbf{n}\,\mbox{d}\Gamma =
     V.&\label{exact_eqn6}
  \end{align}
  \label{exact_eqn}%
\end{subequations}
Here $\Omega$ denotes the two-dimensional fluid domain (the shaded
region in figure~\ref{fig:nondimdomain}(a)) and $\Gamma$ the bubble
boundary,
$p_b(t)$ is the unknown pressure inside the bubble,  $\textbf{R} =
(x_b,y_b)$ is the position of the bubble interface, 
$\kappa$ is the curvature of this 2D interface, and
$\mathbf{n}$ is a unit normal vector directed out of the
bubble. Equation \eqref{exact_eqn1} is the equation for the pressure
in the fluid domain whilst equations \eqref{exact_eqn2} and
\eqref{exact_eqn3} are the dynamic and kinematic conditions on the
bubble surface. Our depth-averaged equations are second order in
space, and hence we can apply only no-penetration conditions,
\eqref{exact_eqn4}; likewise the condition \eqref{exact_eqn2} derives
from the normal component of the dynamic boundary condition at the
bubble interface, and we neglect the tangential stress
balance. \JackRed{Equation \eqref{inlet_equation} imposes the pressure to
  be zero far behind the bubble at $x=-L$ and equation
  \eqref{exact_eqn5} achieves the constant volume flux at $x=L$ by
  imposing a constant pressure gradient.} Additionally, the unknown
value of $p_b$ is determined by ensuring the volume constraint,
\eqref{exact_eqn6}, is satisfied. The computational domain is truncated at a fixed
distance $L$ ahead of and behind the bubble. Following previous
papers, we model the depth-perturbation by a smoothed profile:
\bea b(y) = 1 - \frac{1}{2}h\left[\tanh(s(y + w)) - \tanh(s(y -
  w))\right].
\label{occlusion}
\eea A typical example of this smoothed profile is shown in figure
\ref{fig:nondimdomain}(b).
The dimensionless depth-perturbation height $h$ can be viewed as a
controlled, axially uniform perturbation of the rectangular channel.
The topography profile $b(y)$ enters the equations through the bulk
equation (which is no longer amenable to conformal mapping), the
kinematic boundary conditions and the variable transverse curvature of
the interface.
The unknowns of the problem are $[p,\textbf{R},p_b,U_b]$ and the
control parameters are $[Q,V,h,w,s,\alpha]$. An additional solution measure is the centroid of the bubble, denoted $\overline{y}$, which is useful in determining the symmetry of the system. 
\eqref{exact_eqn} are solved by a finite-element discretisation,
utilising the open-source software \texttt{oomph-lib}
\cite{heil2006oomph}. Details of the implementation can be found  in
\cite{thompson2014multiple} for the case of a propagating air-finger
and \cite{franco2017propagation} for a finite air bubble.

It is convenient to keep $h,w$ and $\alpha$ constant and analyse the
system mainly through variations in $Q$ and $V$, which are easily
manipulated in a physical, experimental set-up. Unless otherwise
specified the results presented in this paper are compared to the
values used by Franco-G{\'o}mez~\textit{et
  al}~\cite{franco2017propagation} and \JackRed{hence a bubble of fixed
  volume $V$ is chosen so that the projected area is $A = \pi r^2$
  with $r=0.46$,} together with a perturbation height $h=0.024$, width
$w=0.25$, sharpness $s=10$ and aspect ratio $\alpha = 40$. 

\section{Initial-value calculations}\label{sec:one}
\setcounter{footnote}{-1}
\begin{figure}
  \centering
  \includegraphics[scale=0.35]{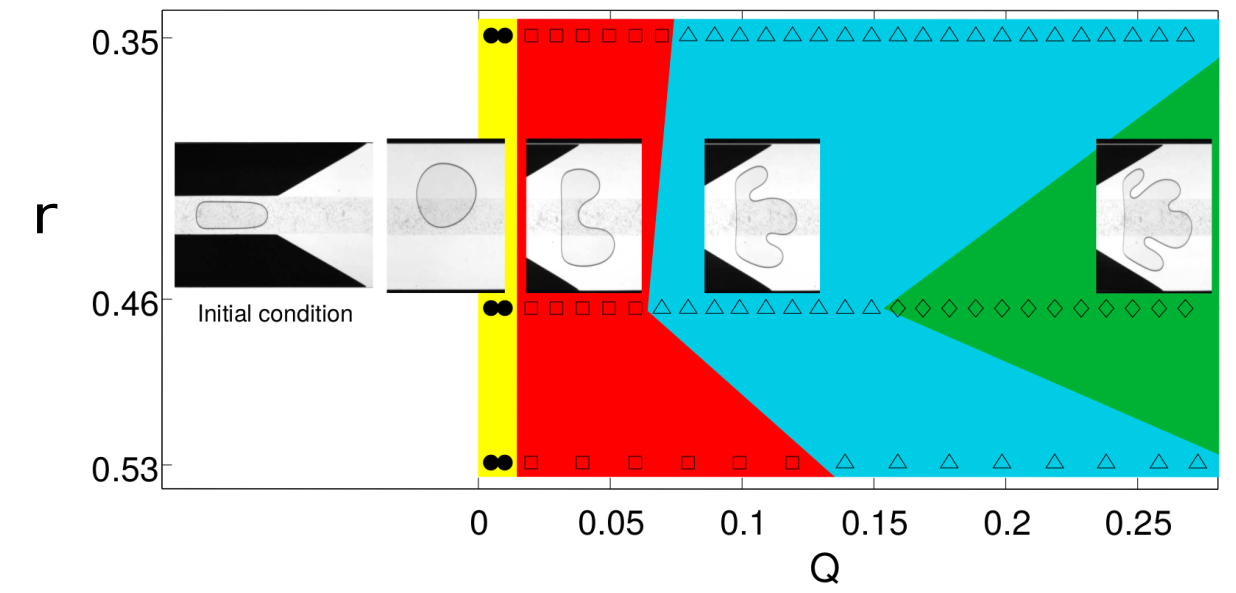}
  \caption[figure4]{\small{Figure adapted with permission from Figure 10 of
    \cite{franco2017propagation}, classifying the fate of circular
    bubbles with static diameter, $2r$ and non dimensional flow-rate
    $Q=\mu U_0^*/\gamma$ according to whether they develop one, two,
    three or four tips before the bubble has either broken up (hollow
    symbols) or remains in a steady mode of propagation (filled
    symbols, only at the lowest flow rate). This data is from
    experiments with $\alpha = 40$, $h = 0.024$ and $w = 0.25$.} \protect\footnotemark}
  \label{fig:fourexp}
\end{figure}
\footnotetext{\copyright The Japan Society of Fluid Mechanics.  Reproduced by permission of IOP Publishing.  All rights reserved.   DOI: 10.1088/1873-7005/aaa5cf}

\begin{figure}
  \centering \includegraphics[scale=0.85]{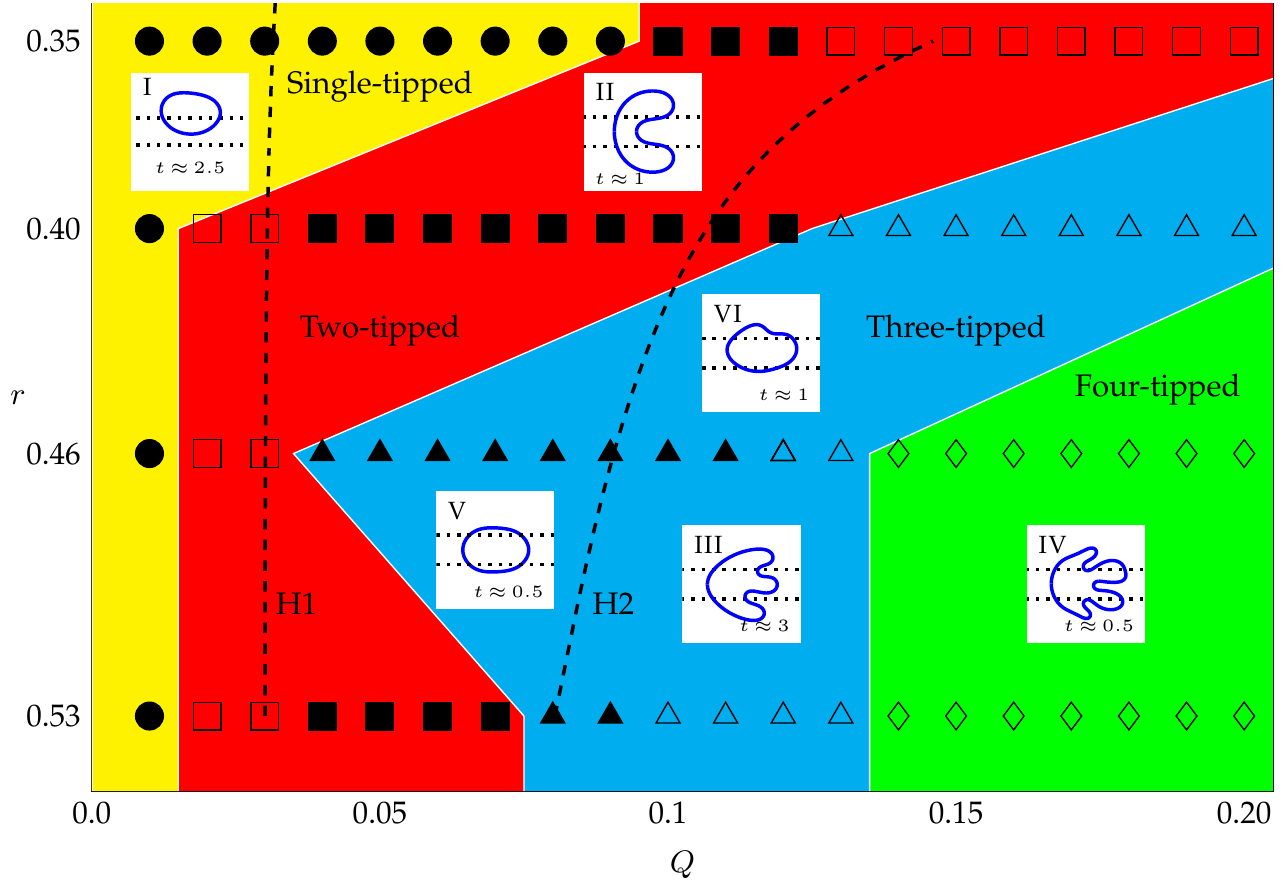}
  \caption{\small{Time-dependent numerical simulations starting from a
      circular bubble with radius (in the $xy$ plane) $r$ and a fixed
      small offset, ${y}_c=0.01$. 
  Solid symbols indicate that the bubble evolves towards either an
  asymmetric steady state, symmetric steady state or stable
  oscillatory motion, labelled by \soltype{1}, \soltype{5} and
  \soltype{6} respectively, while hollow symbols indicate bubble
  breakup.  In the yellow region, the bubble always tend towards a
  single-tipped, steady propagation mode.  In the red region, for
  small flow rates the bubble forms two tips (\soltype{2})
  and breaks up (hollow symbols) or reaches a symmetric steady state
  (solid symbols).  In the blue region, the bubble forms three tips
  before break up (\soltype{3}) or reaching either a
  symmetric steady state or a stable oscillatory state. Finally, in
  the green region, the initial transients involve four-tipped bubbles
  before breaking up (\soltype{4}).  The time quoted in the
  profiles is an indicative dimensionless time-scale at which the
  behaviour in that particular region is typically observed. The
  dashed lines indicate the location of the two Hopf bifurcations in
  the steady solution space that are discussed in
  \S~\ref{sec:two}.}}
\label{fig:four}
\end{figure}
\JackRed{Franco-G{\'o}mez~\textit{et al}~\cite{franco2017propagation}
  performed a range of experiments which showed that an initially
  circular bubble would settle to a stable state below a critical flow
  rate whilst above this threshold the bubble becomes increasingly
  deformed, resulting in a variety of transient outcomes, see
  figure~\ref{fig:fourexp}, including
  topological break up, despite evidence to suggest that an asymmetric
  stable state existed for all flow rates. }

In order to investigate the potential range of evolution in our model
we perform a number of numerical simulations designed to mimic these
experiments. The nonlinear dynamics can be examined by recording the
bubble shape as it propagates along the channel. We do this by solving
the equations in \eqref{exact_eqn} 
{ with the initial condition of a circular bubble with
  radius $r$ centered at $(x_c,y_c) = (0.0,0.01)$ and released from
  rest}.
%
Our simulations reveal a number of qualitatively distinct modes of
propagation. The eventual propagation mode selected has a dependence
on the flow rate $Q$ {and the radius $r$ which sets the
  bubble volume $V$}.
A summary of the qualitative behaviour is shown in
figure~\ref{fig:four},  with regions of parameter space classified
according to the characteristic shape first adopted by the bubble. As
$Q$ and $V$ are varied, we find that the first emergent shape
may have between one and four tips. The start of each different coloured region as $Q$ increases indicates the place in parameter space where the bubble first develops two, three or four
tips for a given bubble volume. {Single tipped bubbles}
evolve smoothly to an asymmetric steadily-propagating configuration
(see inset labelled \soltype{1} in figure~\ref{fig:four}), while
multiple-tipped bubbles may self-intersect in the simulations and thus
break up into one or more fragments, in which case the simulations are
terminated (see insets labelled \soltype{2},\soltype{3},\soltype{4} in
figure~\ref{fig:four}) . In some regimes, the bubble may exhibit
oscillatory behaviour before decaying to either a steady state (see
inset \soltype{5} in figure~\ref{fig:four}) or stable periodic oscillations 
(see inset \soltype{6}). The nature of the oscillatory modes
that tend to a stable state are dependent on the flow rate in that for
lower flow rates  ($Q\approx 0.05$) the bubble appears to oscillate
around a symmetric steady state whilst for larger flow rates
($Q\approx 0.1$) the oscillations manifest  as `waves' above one
edge of the depth-perturbation with the opposite side of the asymmetric
bubble above the other edge of the depth-perturbation as seen in
the \soltype{3} inset in figure~\ref{fig:four}.

We can compare these numerical results to the experimental data of
Franco-G{\'o}mez~\textit{et al}~\cite{franco2017propagation} shown in
figure~\ref{fig:fourexp}. Although the boundaries between regions are
quantitatively different in figures~\ref{fig:fourexp} and
\ref{fig:four}, the phenomenological behaviour is in agreement, and
the transition between different modes of propagation occur over
similar ranges of $Q$ and $V$. For nearly all values of $Q$ explored
in figure~\ref{fig:four}, varying the bubble volume at fixed $Q$
allows us to access a range of qualitatively different modes of
propagation. 

As demonstrated by these numerical time simulations, there is a large
range of possible outcomes from {an initially circular
  bubble} including oscillatory transients and multiple stable steady
states. To understand this further we now perform a detailed analysis
of the steady state bifurcation structure.

\section{Bifurcation Structure for Steady States}\label{sec:two}

\begin{figure}
  \centering
    \includegraphics[width=\textwidth]{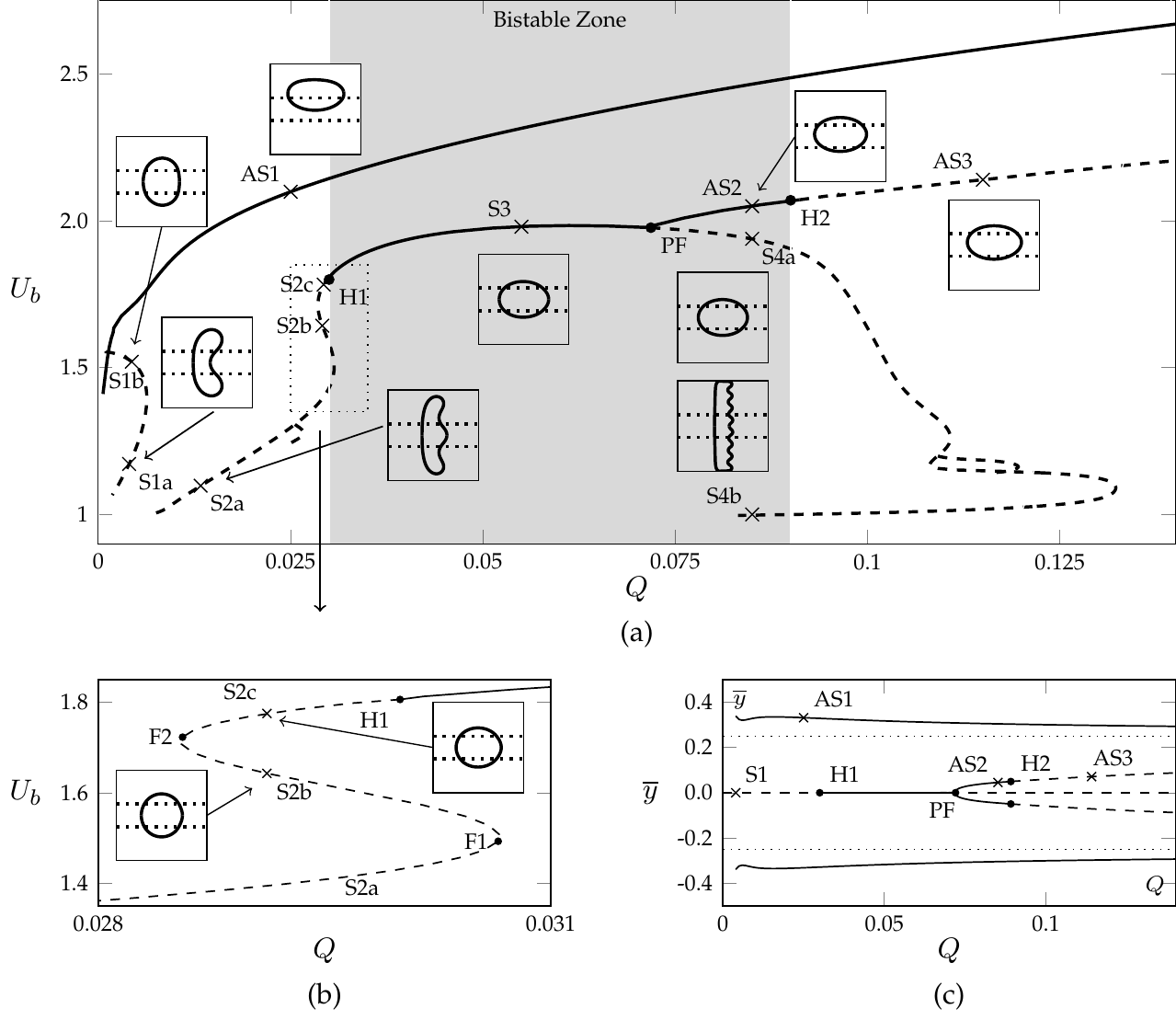}
%
\caption{\small{(a) The steady solution space in the $(Q,U_b)$
    projection for a bubble of volume $V = 0.665$. 
  The solid curves indicate stable solutions whilst the dashed lines
  indicate unstable solutions. Typical shapes on the solution branches
  are labelled by crosses and the pitchfork and two Hopf bifurcations are denoted by
  solid circles. The inset diagrams are bubble shapes on each branch
  that correspond to a cross.  They are on a 1:1 scale in the range
  $x,y\in[-1,1]$. S1a,b solutions $Q = 0.004$, AS1 solution $Q =
  0.025$, S2a,b,c solutions, $Q = 0.029$,  S3 solution $Q = 0.055$,
  AS2 solution $Q = 0.085$, S4a,b solutions $Q = 0.085$, AS3 solution
  $Q=0.115$. The region contained in the dotted rectangle is enlarged
  and shown in panel (b). (c) The steady solution space in the
  $(Q,\overline{y})$ solution space.  }}  \label{fig:b_diag_nonlinear}
\end{figure}

In this section we analyse the steady solution space. We use the flow
rate, $Q$, as a continuation parameter for a fixed bubble volume and
record the bubble speed, $U_b$, and the centroid, $\overline{y}$ as
convenient solution measures.  We also analyse the linear stability of
the numerical solutions as the solution branches are traced out.
%
%
The solution space is shown in figure~\ref{fig:b_diag_nonlinear} in
the $(Q,U_b)$ and $(Q,\overline{y})$ projections respectively, with
stable (unstable) solution branches denoted by solid (dashed)
lines. We note that the branches in this steady solution space are
connected through several bifurcations, in contrast to the $h=0$ case
(see figure~\ref{fig:break_up1}(a)), where the solution branches are
disjoint.


 
For small $Q$ there are three distinct branches as shown in
figure~\ref{fig:b_diag_nonlinear}. The `upper' branch (with largest
$U_b$), AS1, is stable, characterised by an asymmetric bubble shape
and persists for all values of $Q$ that were sampled. The other two
branches, denoted as S1a and S1b, correspond to symmetric
double-tipped solutions. As the value of $h$ is decreased to zero the
S1a branch approaches the $m=0$ solution sketched in
figure~\ref{fig:break_up1}(b) whilst the lower part becomes the
double-tipped solution $m=1$ solution. The S1 branch only exists for a
small range of the parameter $Q$, in contrast to the case with no
depth-perturbation where the solution persists for all $Q$. Both parts
of the S1 branch are unstable; the segment with larger $U_b$ has one
unstable eigenmode, while the slower segment is doubly unstable. 
%
Note that although it appears from
figure~\ref{fig:b_diag_nonlinear}(a) that the S1 and AS1 branches
intersect at small $Q$, 
these branches are actually disjoint as can be seen by the projection
of the solution onto the $(Q,\overline{y})$ plane in
figure~\ref{fig:b_diag_nonlinear}(c).

For $Q\gtrapprox 0.01$, we are able to compute a three-tipped
symmetric solution, S2a, which is unstable. As $Q$ increases this
branch undergoes two fold bifurcations. \JackRed{The first fold, F1,
  (shown in figure~\ref{fig:b_diag_nonlinear}(b)) separates the S2a
  and S2b branches and the second fold, F2, separates the S2b and S2c
  branches}. By decreasing $h$ to zero, the S2a branch converges to a
triple-tipped solution branch, the S2b branch becomes the
double-tipped solution branch and the S2c branch becomes the stable
solution branch (this transition sequence is likely to involve
interaction between the AS1, S1 and S2 branches).  All of the S2
branches are unstable but for this bubble volume and channel geometry,
the solution finally stabilises via a Hopf bifurcation H1 just beyond
the second fold point.
The stable branch of steady solutions that results from this Hopf
bifurcation is labelled S3.

As $Q$ increases further on the S3 branch, a supercritical pitchfork
bifurcation (PF) occurs where the S3 solution breaks symmetry and two
stable asymmetric branches, denoted AS2, emanate from the bifurcation,
see figure~\ref{fig:b_diag_nonlinear}. Beyond the pitchfork, the
unstable symmetric branch experiences multiple bifurcations as $Q$ is
varied and becomes increasingly unstable. This collection of unstable
branches is denoted by S4. {The most extreme shapes on the
  S4 branch, with $U_b$ close to 1, correspond to} the eight-tipped
solution as seen in figure~\ref{fig:b_diag_nonlinear}. Note the
loops on the S2 branch around $Q \approx 0.11$ where an asymmetric
branch connects to the main S4 branch via two pitchfork
bifurcations. This is reminiscent of the so called `snakes and
ladders' bifurcations seen in other nonlinear systems, see, for
example, \cite{burke2007snakes,schneider2010snakes}. This branch has
at least two unstable eigenmodes and as we do not believe it affects
the transient behaviour of the system we do not pursue the details of
this branch here. 
For the largest $Q$ in our calculations ($Q = 0.15$), two distinct asymmetric
states persist in the $(Q,U_b)$ projection, (four in the $(Q,\overline{y})$ projection). The AS1 branch is always stable but the stable AS2
branch experiences a Hopf bifurcation, denoted by H2, and becomes
unstable, the resulting branch labelled AS3. 

In contrast with the $h=0$ solution space, where only one solution is
ever stable, in our case there is a finite-width region of bistability
(with respect to steady states) between the H1 and H2
bifurcations. The AS1 solution is always stable but the nature of the
second stable solution changes from symmetric (the S2 branch) to
asymmetric (the AS2 branch), via the pitchfork
bifurcation. \JackRed{The symmetric stable solution (S2) has been
  observed before in this system, see Franco-G{\'o}mez~\textit{et
    al}~\cite{franco2017propagation}, but its transition to an
  asymmetric stable solution as $Q$ increases, and indeed the presence
  of the two Hopf bifurcations, is a new observation.}
%

\begin{figure}
  \centering \includegraphics[scale=0.85]{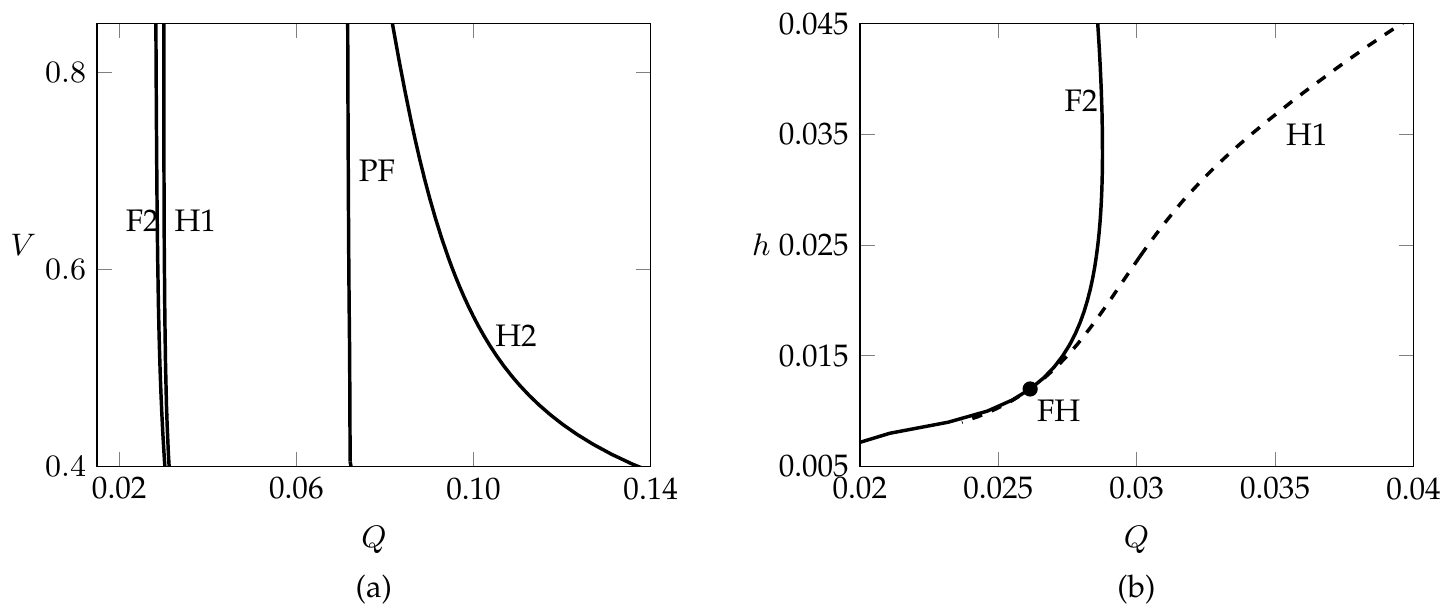}
  \caption{\small{{(a) The location of the bifurcations in the $(Q,V)$
        projection with fixed $h=0.024$. F2 indicates the
        fold bifurcation that occurs before the Hopf bifurcation H1 and PF indicates the pitchfork bifurcation just before H2. (b) The location of the bifurcations in the $(Q,h)$
        projection with $V$ fixed at $V=0.665$. The FH point indicates
        a fold-Hopf bifurcation.}}}
  \label{fig:loci_volume}
\end{figure}

The results presented here are for a fixed volume, $V$, and fixed
depth-perturbation height, $h$ but we can use bifurcation tracking
calculations to investigate the robustness of the location and order
of the bifurcations F2, H1, PF and H2 that bound the bistable
region. The locations of these bifurcations as functions of $V$ and
$h$ are shown in figure~\ref{fig:loci_volume}. We find that the
ordering of the F2, H1, PF and H2 bifurcations are very robust to
changes in volume and, with the exception of H2, their position is
fixed, see figure~\ref{fig:loci_volume}(a). As $V$ decreases, H2
migrates to larger values of $Q$ and hence the width of the bistable
region increases. All the bifurcations are more sensitive to $h$ than
$V$, and when $h<h_c \approx 0.012$,  the change in stability occurs
at the fold, F2, instead of the Hopf, H1, see
figure~\ref{fig:loci_volume}(b).  A fold-Hopf bifurcation occurs when
$h=h_c$.
The dynamics of the system near this type of co-dimension two
bifurcation are very complex and can lead to the appearance of
invariant tori and \JackRed{heteroclinic orbits}, see, for example,
\cite{kuznetsov2013elements}. 

The time-dependent behaviour explored in \S~\ref{sec:one} can now
be interpreted in terms of the bifurcation structure. For all flow
rates, there is an asymmetric stable mode of propagation, which
corresponds to the AS1 branch. The region of bistability appears to
coincide with the solid symbols in figure~\ref{fig:four} which denote
that the bubble is evolving towards a steady solution. The
multi-tipped modes of propagation may relate to the presence of the
unstable S1, S2 and S4 branches, while the oscillating modes of
propagation may be expected to arise due to the Hopf
bifurcations. However, the linear stability analysis does not reveal
the {centre}, nor the stability and size of the periodic
orbits emanating from the Hopf bifurcations and thus the criticality
of H1 and H2 remains unknown. 

There are { at least} three possible approaches to
determine the periodic orbits.  If the periodic orbit is stable, we
expect that for suitable initial conditions, the periodic orbit is
attracting and hence initial-value simulations will eventually
converge towards the periodic orbit. These simulations may be
expensive if the convergence rate is slow, and in any case will not
capture unstable periodic orbits. 
The second approach is to calculate the periodic orbits directly by
solving an extended system of equations as proposed by, for example,
\cite{net2015continuation}. This direct solution will capture unstable
periodic orbits, but leads to a significantly larger system of
equations and requires a discretization in both space and time. The
third approach is to perform a numerical weakly nonlinear stability
analysis near the Hopf points, leading to an analytic  normal form of
the perturbation equations near the Hopf points with numerically
computed coefficients, providing an approximate expression for the
periodic orbits in terms of eigenfunctions.  This latter approach is
the least computationally expensive
and has the advantage of providing semi-analytic approximations for
the amplitude, period and location of the periodic orbits and we
pursue this analysis in the next section.

\section{Weakly Nonlinear Stability}\label{sec:four}

The aim of this section is to perform a local analysis near the Hopf
bifurcation points to obtain approximations for the location,
stability and size of the periodic orbits. In linear stability theory
the growth of the perturbation occurs on a single time-scale and is
determined by the eigenvalues alone and not the amplitude of the
perturbation. In weakly nonlinear stability theory near a marginally
stable solution (i.e. bifurcation point) the growth/decay of a
perturbation is assumed to happen over two time scales, the eigenvalue
providing the growth/decay on the fast scale $t$, whilst the amplitude
varies on a slower time-scale. The \JackRed{main} objective of a
weakly nonlinear analysis is to obtain an evolution equation for this
unknown amplitude function. \JackRed{This method has been applied to a
  number of specific examples in physics, including shear flows,
  shallow water waves, thermoacoustics and magnetohydrodynamics, see,
  for example
  \cite{fujimura1989equivalence,bera_poiseuille,orchiniweaklynonlinear,sanchez2006amplitude,ghidaouiwaterwavenonlinear,laroze2010amplitude,laroze2009amplitude}.}
\JackRed{In these examples the nature of the equations analysed means
  that the algebra often becomes very complicated.} Our approach
(inspired by \cite{sanchez2006amplitude}) is to perform the analysis
for a general set of equations that can be applied to a wide range of
systems so that an analytic approximation for the periodic orbits and
steady states can be obtained. The analysis is based on a continuous
set of equations that are independent of the nature of a
discretisation procedure. We then describe a \JackRed{versatile}
numerical procedure that can be implemented to obtain semi-analytic
expressions for the periodic orbits and steady states. We examine a
set of equations of the form  \bea \mathcal{R}(\dot{{u}},{u},\beta) =
0,
\label{eq0}
\eea where $\mathcal{R}$ is a nonlinear \JackRed{function} that
depends on the state variables $u\in \mathcal{U}$, where $\mathcal{U}$
is an appropriately defined Hilbert space, time derivatives $\dot{u}$
and a parameter $\beta\in\mathbb{R}$. The state variable, $u$, will in
general depend continuously on spatial coordinates, $x$ and temporal
coordinates $t$. To proceed further it is assumed that the set of
equations can be separated so that  \bea
\mathcal{R}(\dot{{u}},{u},\beta) \equiv \mathcal{M}[{u}]\dot{{u}} +
\beta\mathcal{F}[{u}] + \mathcal{G}[{u}]= 0,
\label{general}
\eea where $\mathcal{M}$ is a linear mass operator and $\mathcal{F}$,
$\mathcal{G}$ are nonlinear operators on the state variables
\JackRed{independent of the parameter $\beta$}. The form of the
equations in \eqref{general}, whilst not completely general, are
representative of a large number of physical systems where time
derivatives appear in linear combinations, including the equations that form the subject
of this paper, \eqref{exact_eqn}.

We are interested in the nonlinear evolution of a perturbation near a
Hopf point. A standard linearisation procedure about a steady
solution, $u_s$, will yield a generalised eigenproblem that can be
solved to find the eigenmodes, denoted $g$, and corresponding
eigenvalues, denoted $s$. A Hopf point occurs at a particular value of
the parameter, $\beta_c$ say, such that a single pair of eigenvalues
are located on the imaginary axis, i.e. of the form
$s=\pm\mathrm{i}\omega_c$. Near this point of marginal stability the
oscillations will occur on a fast time scale and will be modulated by
a time-dependent amplitude function that operates on a slower time
scale. Following
\cite{sanchez2006amplitude,orchiniweaklynonlinear,drazin2004hydrodynamic}
we define a fast time scale, $t_0=t$ and a slow time scale, $t_1 =
\varepsilon^2 t$, where $\varepsilon\ll 1$ is a small unfolding
parameter, and employ the method of multiple scales. The solution,
$u$, is now a function of two different time scales as well as the
spatial variables denoted by $x$. The solution, time derivative and
parameter $\beta$ are therefore expanded in ascending powers of
$\varepsilon$:
\begin{subequations}
  \begin{align}
  {u} =&\: {u}_s(x) + \varepsilon{u}_1(t_0,t_1,x) +
  \varepsilon^2{u}_2(t_0,t_1,x) +
  \varepsilon^3{u}_3(t_0,t_1,x)+\mathcal{O}(\varepsilon^4),\label{asym1}\\ \pdiff{}{t}
  =&\: \pdiff{}{t_0} + \varepsilon^2\pdiff{}{t_1} +
  \mathcal{O}(\varepsilon^4), \label{asym2}\\ \beta =&\:\beta_c(1 +
  \delta\varepsilon^2).\label{asym3}
  \end{align}
  \label{asym}%
  \end{subequations}
The parameter $\delta = \pm 1$ signifies which `side' of the Hopf
bifurcation is being analysed. Substituting the expressions in
\eqref{asym} into \eqref{general} yields a sequence of linear problems
that can be solved at each order of $\varepsilon$. 

Examining terms of $\mathcal{O}(\varepsilon^0)$ in the expansion gives
the following equation to solve for the steady base solution, \bea
\beta_c\mathcal{F}[{u}_s] + \mathcal{G}[{u}_s]= 0.
\label{pence}
\eea \JackRed{With $u_s$ known}, we continue further and at
$\mathcal{O}(\varepsilon)$ we have \bea
\mathcal{L}\left(\pdiff{{u}_1}{t_0},{u}_1\right) \equiv
\mathcal{M}[{u}_s]\pdiff{{u}_1}{t_0} + \mathcal{J}[{u}_s]{u}_1 = 0,
\label{jack}
\eea where the operator $\mathcal{J}[u_s]$ is the Jacobian operator
defined by its action on $v_1\in \mathcal{U}$ by the Fr\'{e}chet
derivative, \JackRed{\begin{equation} \mathcal{J}[{u_s}]v_1 \equiv
    D_{v_1}\mathcal{K}[u_s] \equiv \lim_{k\to
      0}\frac{\mathcal{K}[u_s+kv_1] - \mathcal{K}[u_s]}{k},
    \label{frechjaco}
\end{equation}}
\JackRed{where $\mathcal{K}[u]\equiv\beta_c\mathcal{F}[u] +
  \mathcal{G}[u]$. }\JackRed{We separate $u_1$ into spatial and
  temporal parts so that }$u_1(t_0,t_1,x) = A(t_1)g(x)\mbox{e}^{st_0}$
and equation \eqref{jack} becomes a generalised eigenproblem. Its
solution determines the linear stability of the steady solutions
solved in \eqref{pence}. \JackRed{There are an infinite number of
  solutions to \eqref{jack} but as we are only interested in periodic
  solutions arising from a Hopf bifurcation, we choose $u_1$ to be}
\bea {u}_1 = A(t_1)\,{g}(x)\,\mbox{e}^{\mathrm{i}\omega_c
  t_0}+\mbox{c. c.},
\label{homog}
\eea where $A(t_1)$ is the undetermined amplitude of the perturbation
depending on the slow time scale $t_1$. The function ${g}(x)$ is a
complex eigenfunction of the generalised eigenvalue problem,
$\mathrm{i}\omega_c\mathcal{M}{g} = \mathcal{J}{g}$. Proceeding to
$\mathcal{O}(\varepsilon^2)$ the problem to be solved is \bea
\mathcal{L}\left(\pdiff{{u}_2}{t_0},{u}_2\right) =
        {-\mathcal{M}[{u}_1]\pdiff{{u}_1}{t_0}   - \delta
          \beta_c\mathcal{F}[{u}_s]-\frac{1}{2}\mathcal{H}[{u}_s]({u}_1,{u}_1)},
\label{secondorder}
\eea \JackRed{where $\mathcal{H}[u_s]$ represents the bilinear Hessian
  operator defined its action on $v_1,v_2\in \mathcal{U}$ and using
  \eqref{frechjaco} by
  \begin{align}
    \mathcal{H}[{u_s}](v_1,v_2)  \equiv
    D_{v_2}(\mathcal{J}[u_s]v_1)\equiv\lim_{k\to
      0}\frac{\mathcal{J}[u_s + kv_2]v_1 - \mathcal{J}[u_s]v_1}{k}.
    \label{frechhess}%
  \end{align}}The terms on the right hand side { of \eqref{secondorder}} are either formed of products of $u_1$ or have no time dependence. It can be shown that the products of $u_1$, defined in \eqref{homog}, introduce time dependent terms proportional to $\mbox{exp}(2\mathrm{i}\omega_ct_0)$ { in \eqref{secondorder}} and these do not resonate with the homogeneous solutions of the operator $\mathcal{L}$. By examining the form of ${u}_1$ in \eqref{homog} a particular solution to \eqref{secondorder} is sought in the form
\begin{equation}
  {u}_2 =  A^2{\varphi}_0\,\mbox{e}^{2\mathrm{i}\omega_c t_0} +
  \mbox{c. c.} + |A|^2{\varphi}_1+{\varphi}_2,
 \label{form}%
\end{equation}
where the functions $\varphi_i$ are to be determined. Substituting
\eqref{form} into \eqref{secondorder} and then equating coefficients
of $A^2$, $|A|^2$ and constant terms leads to three linear equations
that can be solved to find the undetermined functions ${\varphi}_i$ in
\eqref{form}. These equations can be stated as
\begin{subequations}
  \begin{align}
&  \left(2\mathrm{i}\omega_c\mathcal{M}[{u}_s] +
    \mathcal{J}[{u}_s]\right){\varphi}_0
    =\:-\mathrm{i}\omega_c\mathcal{M}[{g}]{g}-\frac{1}{2}\mathcal{H}[{u}_s]({g},{g}),\\  &\mathcal{J}[{u}_s]{\varphi}_1
    =\:
    -\frac{1}{2}\mathcal{H}[{u}_s]({g},{g}^*)-\frac{1}{2}\mathcal{H}[{u}_s]({g}^*,{g})
    - \mathrm{i}\omega_c\mathcal{M}[{g}^*]{g} +
    \mathrm{i}\omega_c\mathcal{M}[{g}]{g}^*,\\ &\mathcal{J}[{u}_s]{\varphi}_2
    = \: - \delta \beta_c\mathcal{F}[{u}_s],
  \end{align}
  \label{east}%
\end{subequations}
where the star denotes complex conjugation. The amplitude function
$A(t_1)$ remains undetermined at this order. Therefore the analysis
continues to the next order, $\mathcal{O}(\varepsilon^3)$. The
corresponding equation to be solved is
\begin{multline}
  \mathcal{L}\left(\pdiff{{u}_3}{t_0},u_3\right) =
          {-\mathcal{M}[{u}_2]\pdiff{{u}_1}{t_0} -
            \mathcal{M}[{u}_1]\pdiff{{u}_2}{t_0} -
            \mathcal{M}[{u}_s]\pdiff{{u}_1}{t_1}}  \\ {-\delta
            \beta_c\mathcal{J}_{\mathcal{F}}[{u}_s]{u}_1}
          {-\frac{1}{2}\mathcal{H}[{u}_s]({u}_1,{u}_2) -
            \frac{1}{2}\mathcal{H}[{u}_s]({u}_2,{u}_1)}  {-
            \frac{1}{6}\mathcal{T}[{u}_s]({u}_1,{u}_1,{u}_1)},
\label{errol}%
\end{multline}
where $\mathcal{J}_{\mathcal{F}}[u_s]u_1\equiv
D_{u_1}\mathcal{F}[u_s]$ and  the operator $\mathcal{T}[u]$ is the
trilinear third-order differential operator defined by its action on
$v_1,v_2,v_3\in \mathcal{U}$ and using \eqref{frechhess} by
\begin{align}
  \mathcal{T}[{u}](v_1,v_2,v_3)  \equiv
  D_{v_3}(\mathcal{H}[u_s](v_1,v_2)) \equiv \lim_{k\to
    0}\frac{\mathcal{H}[u_s+kv_3](v_1,v_2) -
    \mathcal{H}[u_s](v_1,v_2)}{k}.
  \label{frechtress}%
\end{align}
At this order the products formed on the right-hand side of
\eqref{errol} will result in time-dependent terms proportional to
$\mbox{exp}(\mathrm{i}\omega_ct_0)$ and these \textit{are} resonant
with homogeneous solutions of the operator $\mathcal{L}$. Therefore a
solvability condition is invoked. In linear operator theory, by the
Fredholm alternative (see, for example, \cite{boyce1969elementary}), a
linear system of the form $\mathcal{L}u = f$ will either have (a) a
unique solution $u$, or (b) a non-trivial solution to
$\mathcal{L}^{\dagger}v = 0$. For condition (a) to hold $\langle
v^{\dagger}, f\rangle = 0$. The dagger superscripts indicate the
adjoint problem. The adjoint problem and inner product on the Hilbert
space, $U$, are defined by 
$$ \langle \mathcal{L}u,v\rangle  = \langle
u,\mathcal{L}^{\dagger}v\rangle,\qquad \langle u,v \rangle =
\int_{0}^{T}\int_{x\in\Omega} u v^* \,\mbox{d} x\,\mbox{d}t_0
$$ for any functions $u$ and $v$ belonging to the solution function
space, where $\Omega$ is the domain of definition of the spatial
variable $x$, and $T$ is the period of oscillation in the fast time
scale $t_0$. Applying the Fredholm alternative to \eqref{errol} yields
a solvability condition which provides a constraint on the amplitude
function $A(t_1)$. This can be written as \bea \hat{\nu}
\pdiff{A}{t_1}  + \hat{\lambda} A + \hat{\mu} A|A|^2=0,
\label{ls2}
\eea which is the weakly nonlinear Landau equation. The coefficients,
$\hat{\nu}$, $\hat{\lambda}$ and $\hat{\mu}$ are defined by \bea
\hat{\nu} = \langle
\mathcal{M}[{u}_s]{g},{g}^{\dagger}\rangle,\qquad\hat{\lambda}=
\sum_{k=0}^3\langle \Gamma_k,{g}^{\dagger}\rangle,\qquad\hat{\mu} =
\sum_{k=0}^7\langle \Lambda_k,{g}^{\dagger}\rangle,
\label{storm}
\eea where $g^{\dagger}$ is the eigenfunction corresponding to the
adjoint problem, $\mathcal{L}^{\dagger}g^{\dagger} = 0$. The values of
$\Lambda_k$ and $\Gamma_k$ are defined as 
\begin{equation}
  \begin{split}
  \Gamma_0 = \:& \delta \beta_c\mathcal{J}_{\mathcal{F}}[{u}_s]{g},
  \quad \Gamma_1 = \:
  \mathrm{i}\omega_c\mathcal{M}[{\varphi}_2]{g},\quad \Gamma_2 = \:
  \frac{1}{2}\mathcal{H}[{u}_s]({g},{\varphi}_2), \quad \Gamma_3 =\:
  \frac{1}{2}\mathcal{H}[{u}_s]({\varphi}_2,{g}),\\ \Lambda_0 =\:&
  -\mathrm{i}\omega_c\mathcal{M}[{\varphi}_0]{g}^*,\quad\Lambda_1 = \:
  \mathrm{i}\omega_c\mathcal{M}[{\varphi}_1]{g}, \quad\Lambda_2 = \:
  2\mathrm{i}\omega_c\mathcal{M}[{g}^*]{\varphi}_0, \quad\Lambda_3 =
  \: \frac{1}{2}\mathcal{H}[{u}_s]({g},{\varphi}_1),\\\Lambda_4 = \:&
  \frac{1}{2}\mathcal{H}[{u}_s]({g}^*,{\varphi}_0), \quad\Lambda_5 =
  \: \frac{1}{2}\mathcal{H}[{u}_s]({\varphi}_1,{g}),\quad \Lambda_6 =
  \frac{1}{2}\mathcal{H}[{u}_s]({\varphi}_0,{g}^*),\\ \Lambda_7 = \:&
  \frac{1}{6}\mathcal{T}[{u}_s]({g}^*,{g},{g})+\frac{1}{6}\mathcal{T}[{u}_s]({g},{g}^*,{g})+\frac{1}{6}\mathcal{T}[{u}_s]({g},{g},{g}^*).
  \end{split}
  \label{grizzly}
\end{equation}
The values of $\hat{\nu},\hat{\lambda},\hat{\mu}$ can be obtained once
the functions $\varphi_i$ and the eigenmodes $g$ and $g^\dagger$ have
been calculated. For convenience the eigenmodes $g$ and $g^\dagger$
are normalised so that $|\langle \mathcal{M}g,g^\dagger\rangle| =
1$. Therefore, $|\nu| = 1$. Furthermore it is more convenient to write
the Landau equation as \bea \pdiff{A}{t_1} = \lambda A + \mu A|A|^2,
\label{ely}
\eea where $\lambda = -\hat{\lambda}/\hat{\nu}$ and $\mu =
-\hat{\mu}/\hat{\nu}$.

The analysis up to this point has been based on a spatially and
temporally continuous set of equations. We now describe a procedure,
independent of the choice of discretisation, appropriate for a
spatially discretised version of the equations in \eqref{eq0}. In the
description of the algorithm that follows, $\textbf{J}$ and
$\textbf{M}$ are the matrix representations of $\mathcal{J}$ and
$\mathcal{M}$ respectively and bold symbols are the discretised
vectors of their continuous counterpart. It is assumed that efficient
numerical linear solvers (for example, SuperLu \cite{li2005overview}),
generalised eigensolvers (for example, Trilinos \cite{herouxtrilnos}).
and continuation algorithms are available. We note that for
high-dimensional systems, these calculations are computationally demanding, but they
need to be performed only once for each bifurcation point.  The
procedure is as follows:
\begin{enumerate}
\item Find a Hopf bifurcation as $\beta$ is varied by a continuation
  method as described by, for example, \cite{kuznetsov2013elements}.
\vspace{0.1cm}

\item Calculate the solution of \eqref{pence}, $\textbf{u}_s$, using
  Newton's method at $\beta = \beta_c$. 
\vspace{0.1cm}

\item Calculate the eigenfunction $\textbf{g}$ and eigenvalue
  $s_c=\mathrm{i}\omega_c$ using \JackRed{an} eigensolver to solve
  $\textbf{J}[\textbf{u}_s]\textbf{g}=s\textbf{M}[\textbf{u}_s]\textbf{g}$.
\vspace{0.1cm}

 \item Calculate the Hessian products on the right hand side of
   \eqref{east}. 
If an exact expression for the Jacobian matrix is available, the
Hessian products can be numerically computed using the
central-difference formula \bea \mathcal{H}[{u}_s](v_1,v_2) \approx
\frac{\textbf{J}[{\textbf{u}}_s+k_1\textbf{v}_2]\textbf{v}_1 -
  \textbf{J}[\textbf{u}_s - k_1\textbf{v}_2]\textbf{v}_1}{2k_1},
    \label{hess_num}
   \eea where $k_1$ is a small finite difference parameter.
\vspace{0.1cm}

\item Calculate the functions
  $\boldsymbol{\varphi}_0,\boldsymbol{\varphi}_1$ and
  $\boldsymbol{\varphi}_2$ in \eqref{east} using a linear solver.
\vspace{0.1cm}

\item Calculate the adjoint eigenfunction, $g^{\dagger}$ by solving
  $\textbf{J}^T[\textbf{u}_s]\textbf{g}^{\dagger}=-s\textbf{M}^T[\textbf{u}_s]\textbf{g}^{\dagger}$.
\vspace{0.1cm}

\item Calculate the Hessian and third-order products in
  \eqref{grizzly} using \eqref{hess_num} by approximating the
  third-order operator (again assuming an exact Jacobian) as    
\begin{multline}\nonumber
    \mathcal{T}[u_0](v_1,v_2,v_3)
    \approx\\ \frac{\textbf{J}[\textbf{u}_s+k_2\textbf{v}_2+k_2\textbf{v}_3]\textbf{v}_1
      - \textbf{J}[\textbf{u}_s + k_2\textbf{v}_2]\textbf{v}_1 -
      \textbf{J}[\textbf{u}_s-k_2\textbf{v}_2+\textbf{v}_3]\textbf{v}_1
      +
      \textbf{J}[\textbf{u}_s-k_2\textbf{v}_2]\textbf{v}_1}{2k_2^2}.\end{multline}
    
    \item Finally combine the quantities in \eqref{grizzly} to
      calculate the values of $\hat{\nu},\hat{\lambda}$ and
      $\hat{\mu}$ given in \eqref{storm}, defining the discrete inner
      product as $\langle \textbf{f},\textbf{g}\rangle =
      \textbf{f}\textbf{g}^{T}$.
\end{enumerate}

\JackRed{We now return our focus to the specific equations in
  \eqref{exact_eqn}}. These are discretised by the finite-element
method, utilising the open-source software \texttt{oomph-lib}
\cite{heil2006oomph} in which the functions required for the above
procedure have all been implemented. It is convenient to choose the
independent parameter $\beta = Q^{-1}$ and, to ensure the finite
difference formulas stated above converge, $k_1=10^{-4}$ and
$k_2=10^{-3}$. When $V = \pi r^2$, with $r=0.46$, $\alpha=40$,
$w=0.25$ and $h=0.024$, this procedure was implemented and the
critical values of $Q_c$, $\omega_c$ and the Landau coefficients were
found to 3 significant figures to be
  \begin{subequations}
    \begin{align}
      \mbox{H}1: Q_c = Q_{1} = 0.0301, \quad \omega_c=3.58,
      \quad&\:\lambda = \delta(5.63 -10.7\mathrm{i}),\quad \mu =
      1280-1200\mathrm{i}, \label{hopf1}\\ \mbox{H}2: Q_c = Q_{2} =
      0.0896, \quad \omega_c=10.8, \quad&\:\lambda = \delta(-5.43 -
      15.8\mathrm{i}),\quad \mu = -6150 -
      1320\mathrm{i}. \label{hopf2}
    \end{align}
    \label{hopf}%
 \end{subequations}
\JackRed{We have checked that the quadrant in the complex plane of
  which $\lambda$ and $\mu$ lie (setting $\delta$ constant) does not
  change as parameters are varied thus not affecting the criticality
  of the bifurcation.}

At this point, we wish to analyse solutions to \eqref{ely}. We note
that a decomposition into modulus argument form, i.e. $A =
\hat{r}(t)\exp(\mathrm{i}\theta)$ converts the equation into a pair of
ODEs:
%
\begin{equation}
  \diff{\hat{r}}{t_1} = a\hat{r} + b\hat{r}^3,\qquad
  \diff{\theta}{t_1} =  c + d\hat{r}^2,
  \label{jaco1}%
\end{equation}
where $a = \delta \Re(\lambda)$, $b = \Re(\mu)$, $c =
\delta\Im(\lambda)$ and $d = \Im(\mu)$.  The invariant solutions occur
when $\mbox{d}{\hat{r}}/\mbox{d}{t_1} = 0$. The solutions
corresponding to $\hat{r}=0$ are the steady states of the fully
nonlinear steady equations, while the  solution corresponding to
$\hat{r}=\sqrt{-a/b}$ is an invariant periodic orbit of the fully
nonlinear system.
%
By examining the signs of $a$ and $b$ at each Hopf point and
performing a stability analysis on the equations in \eqref{jaco1} it
is straightforward to show that H1 is subcritical; unstable periodic
orbits exist when $Q>Q_1$ for $Q$ near to $Q_1$. Similarly, H2 is
supercritical so stable periodic orbits exist when $Q>Q_2$. 

\JackRed{Near the H1 and H2 points the steady state solution of
  \eqref{eq0}, corresponding to $\hat{r}=0$ in \eqref{jaco1}, can be
  written in terms of the state variables as}  \bea u = {u}_s +
\varepsilon^2{\varphi}_2 + \mathcal{O}(\varepsilon^4).
\label{steadyapprox}%
\eea
\begin{figure}
  \centering \includegraphics[scale=0.85]{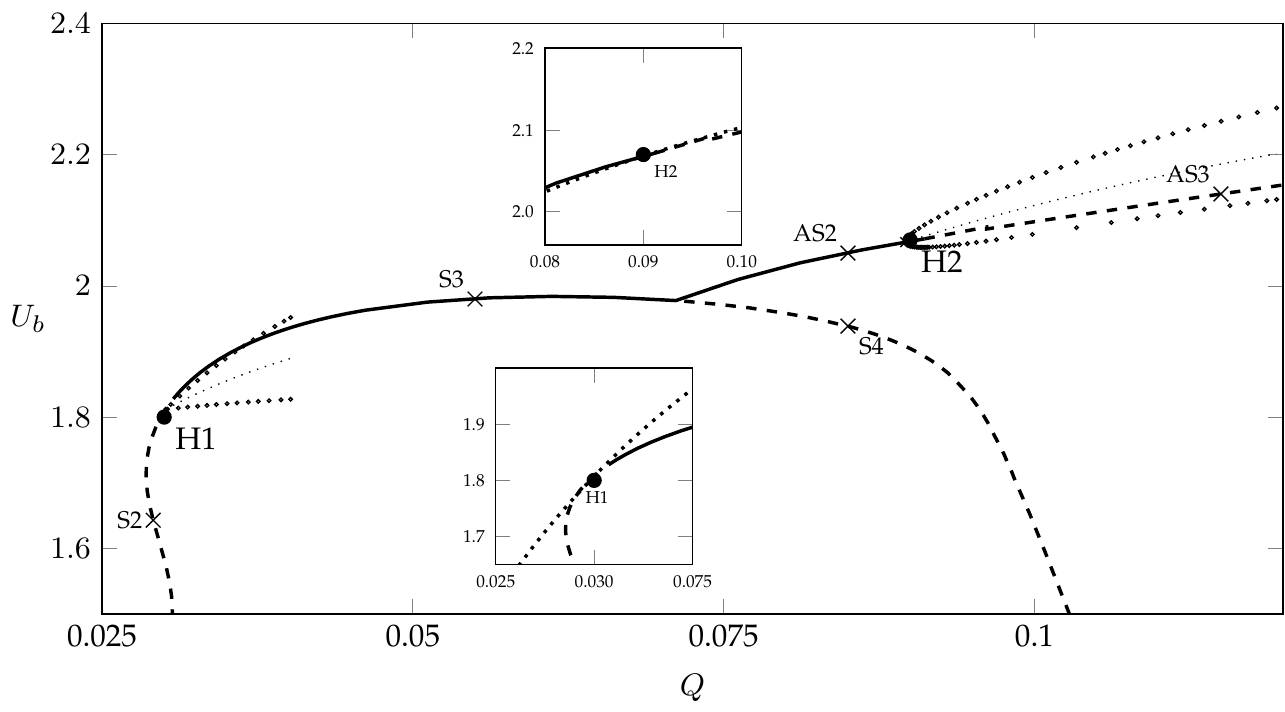}
  \caption{\small{{Comparison of the fully nonlinear solution space in
        the $(Q,U_b)$ plane with the weakly nonlinear
        approximation. The dotted lines indicate the centre of
        oscillation of the periodic orbits and the circular markers
        are the amplitude as given in \eqref{limitapprox}. The solid
        and dashed curves in the inset diagrams \JackRed{(enlargements
          near H1 and H2)} are the fully nonlinear steady solutions,
        whilst the black dotted curves are the weakly nonlinear
        approximations given by \eqref{steadyapprox}.}}}
  \label{fig:b_diag_wnl}%
\end{figure}
This weakly nonlinear approximation can be compared to the fully nonlinear steady state bifurcation structure of \S~\ref{sec:two}. The dotted lines in the inset panels of figure~\ref{fig:b_diag_wnl} are the approximations given by \eqref{steadyapprox}. The approximation near the H2 bifurcation point is excellent \JackRed{in this projection}; the fully nonlinear and weakly nonlinear curves are visually indistinguishable. For the H1 bifurcation point the weakly nonlinear steady state approximation deviates significantly away from the solution branch as we move away from H1 in the $(Q,U_b)$ projection. This difference { is likely due to} the existence of the fold, F2, in the immediate vicinity of H1. As $h$ is varied, to approach the fold-Hopf point, a weakly nonlinear analysis should include the eigenmode associated with the zero eigenvalue and hence the form of the perturbation equation in \eqref{ely} would differ. 

The periodic orbits, corresponding to $\hat{r}=\sqrt{-a/b}$, can be
stated as  \bea {u} = {u}_s + \varepsilon \sqrt{-\frac{a}{b}}
\,{g}\mbox{e}^{\mathrm{i}(\omega_ct_0+\Delta)}+
\varepsilon^2\left({\varphi}_2-\frac{a}{b}{\varphi}_1-\frac{a}{b}{\varphi}_0\mbox{e}^{2\mathrm{i}(\omega_ct_0+\Delta)}\right)
+ \mathcal{O}(\varepsilon^3),
\label{limitapprox}
\eea where $\theta=\theta_0$ at $t=0$ and $\Delta = \theta_0 +
\varepsilon^2(c-ad/b)t_0$ represents the $\mathcal{O}(\varepsilon^2)$
correction to the phase of the oscillations. The location of the
centre of oscillations \JackRed{(ignoring time-dependent terms in
  \eqref{limitapprox})} of these periodic orbits are shown as dotted
lines in the main panel of figure~\ref{fig:b_diag_wnl}. The circular
markers in this diagram indicate the amplitude of the periodic
orbits. The initial phase $\theta_0$ will not alter the amplitude or
period of the periodic orbit.

\begin{figure}
  \centering
  \includegraphics[scale=0.62]{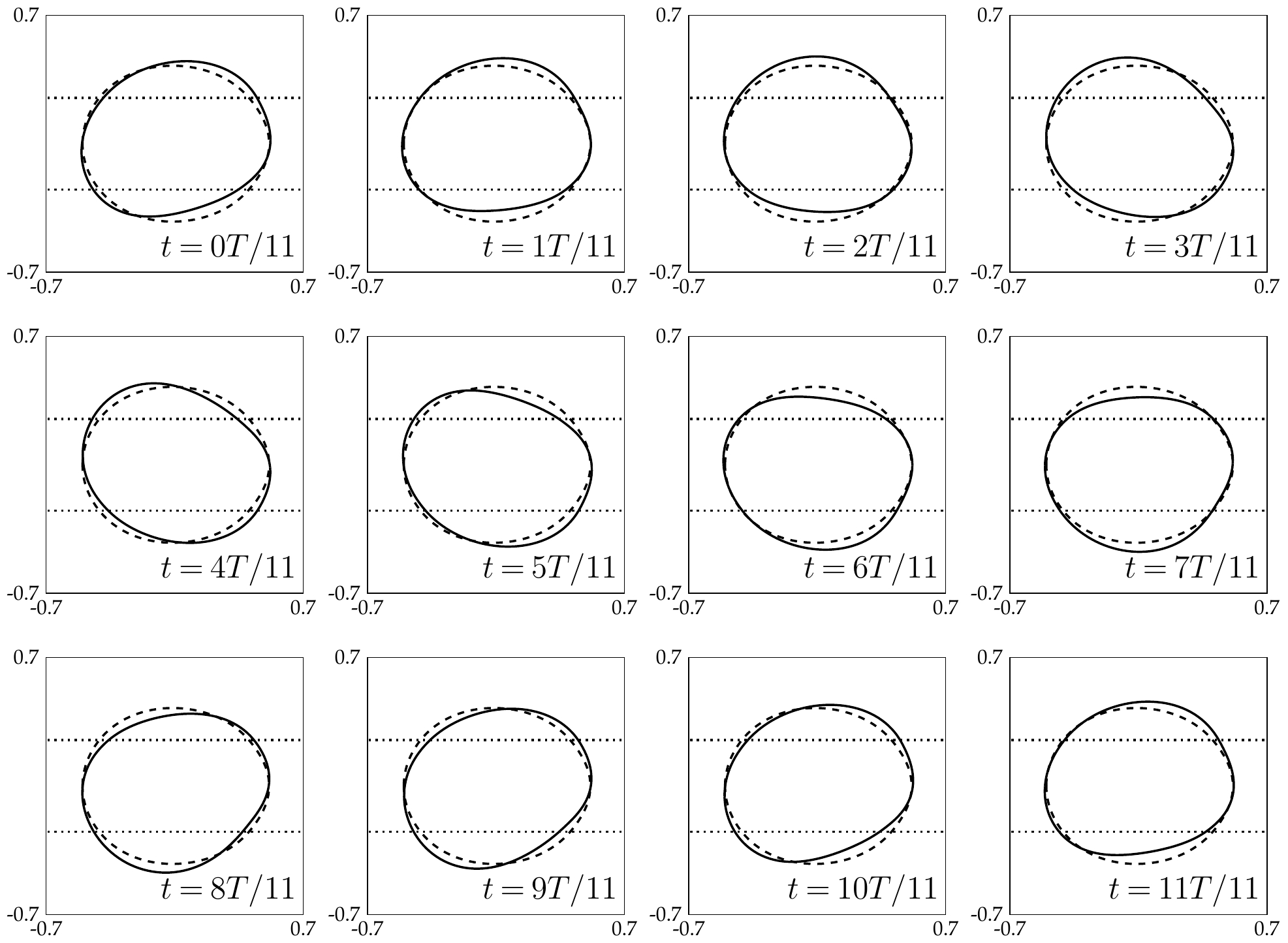}  
  \caption{\small{Unstable weakly periodic orbit near H1. The dashed
      shape is the fully nonlinear steady state solution. Bubble
      shapes for $Q^{-1} = Q_{1}^{-1}(1-\varepsilon^2)$, where $Q_1$
      is the critical value for H1 and $\varepsilon=0.2$. The period
      is given by $T = 2\pi/\omega$ where $\omega = \omega_c +
      \varepsilon^2(c - ad/b)$ which in this case is $T = 1.657$.}}
  \label{fig:periodic_orbits_h1_2}
\end{figure}

An example of the weakly nonlinear bubble shapes of the periodic
orbits near the H1 point is shown in
figure~\ref{fig:periodic_orbits_h1_2}. The oscillations are
characterised by an oscillation about the symmetric steady solution
and the oscillations are visible on both sides of the bubble. It can
be seen that after half a period the bubble has `flipped' from its
initial shape. The weakly nonlinear equations in \eqref{jaco1} predict
that if the initial amplitude, $r_0$, of a perturbation is smaller
than {a critical amplitude} $\hat{r}$, then the system will
evolve towards $r=0$, i.e. the S3 steady state. In contrast if
$r_0>\hat{r}$ then the trajectory will diverge to $r\to\infty$. The H1
periodic orbit can therefore be considered an edge state of the
reduced system as it forms the boundary between stable and unstable
behaviour.

\begin{figure}
  \centering
  \includegraphics[scale=0.94]{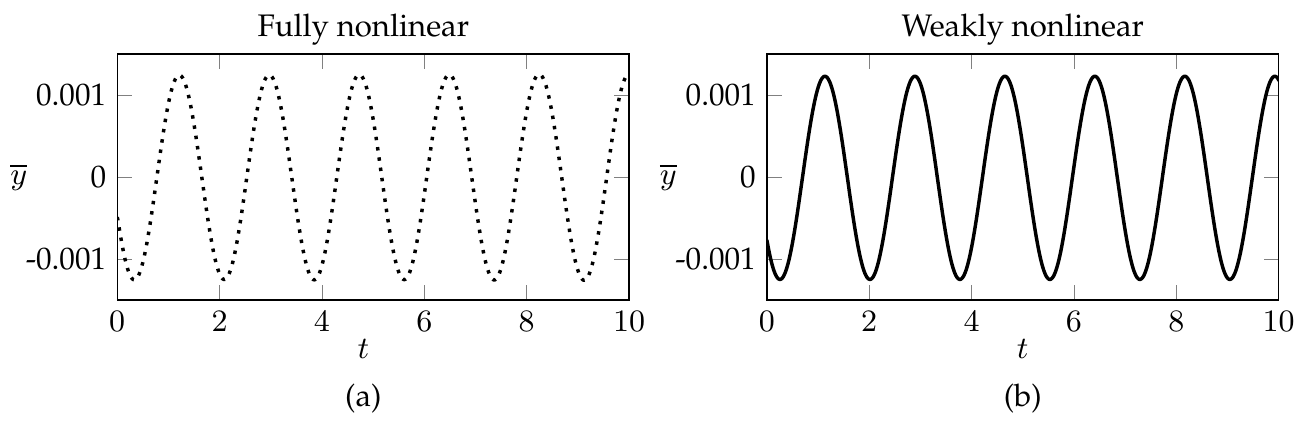}  
  \caption{\small{Comparison between the fully nonlinear
      time-dependent solutions and the weakly nonlinear solutions near
      H1. $Q^{-1}=Q_1^{-1}(1-\varepsilon^2)$ with
      $\varepsilon=0.01$. The dotted curves and solid curves are the
      fully nonlinear and weakly nonlinear solutions respectively. (a)
      and (b) The time signal of $\overline{y}$.}}
  \label{fig:periodic_orbits_h1_1}
\end{figure}

\begin{figure}
  \centering
  \includegraphics[scale=0.94]{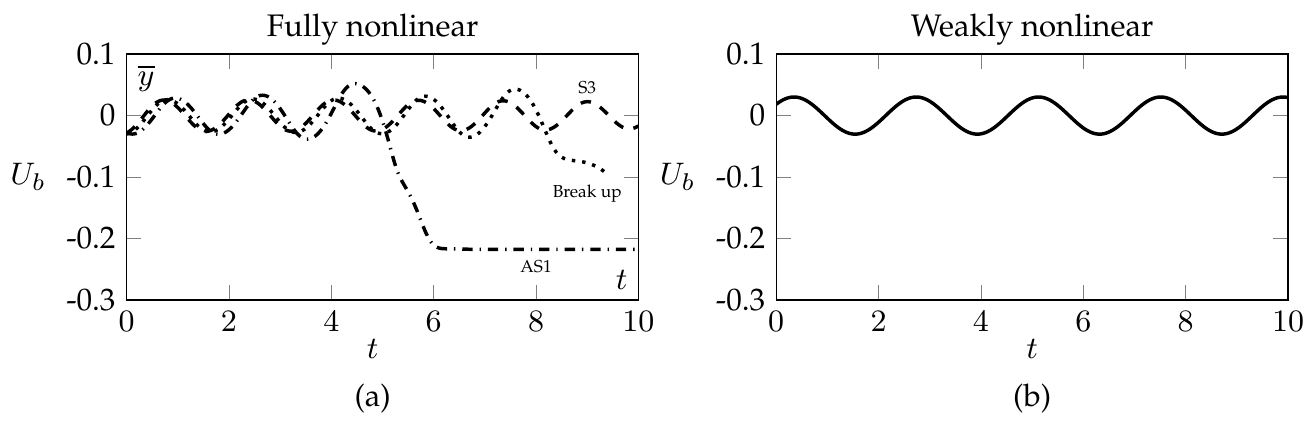}  
  \caption{\small{Comparison between the fully nonlinear
      time-dependent solutions and the weakly nonlinear solutions near
      H1, $Q = 0.032$. (a) The time signal of $\overline{y}$ for the
      fully nonlinear simulations. The dashed line is for
      $\theta_0=\pi/2$. The dotted is for $\theta_0=5\pi/12$ where the
      bubble eventually breaks up and the dashed-dot line is for
      $\theta_0=\pi/4$ when the bubble reaches the AS1 steady state
      (b) The weakly nonlinear simulations when $\theta_0 = 0$.}}
  \label{fig:periodic_orbits_h1_3}
\end{figure}

The orbits near H1 are unstable so as noted in \S~\ref{sec:two} we are
unable to precisely capture the fully nonlinear periodic orbit by
time-integration alone. However by choosing the weakly nonlinear
solution in \eqref{limitapprox} as an initial \JackRed{condition} to
the fully nonlinear time-dependent problem a qualitative comparison
can be made. A characteristic feature of the unstable periodic orbits
is that the perturbation oscillates all the way around the edge of the
bubble (see figure~\ref{fig:periodic_orbits_h1_2}). \JackRed{These
  oscillations will cause oscillations in the centroid of the
  bubble. Therefore a useful way to compare the fully nonlinear
  simulations and weakly nonlinear approximations is to measure the
  time signal of $\overline{y}$ so that the amplitude and period can
  be compared.} A result of one such simulation when
$\varepsilon=0.01$ is shown in figure~\ref{fig:periodic_orbits_h1_1}
in the fully nonlinear simulations (a), and weakly nonlinear
simulations, (b). The period and amplitude of oscillations of both
regimes are nearly identical in both regimes. At this value of
$\epsilon$, we are very close to the H1 point and hence the
oscillations in the fully nonlinear regime will take a long time to
decay. For larger values of $\varepsilon$ we find that depending on
the initial phase of the initial condition, the bubble
will either break up, evolve to the AS1 solution or continue to
oscillate towards the S3 solution, see
figure~\ref{fig:periodic_orbits_h1_3}. This is further evidence that
the H1 orbit is an edge state in the fully nonlinear regime. In
contrast to the weakly nonlinear regime there are three possible
outcomes with the fully nonlinear H1 periodic edge state forming the
dividing line between stable symmetric evolution and either breaking
up or being attracted towards the stable asymmetric states.

\begin{figure}
  \centering
  \includegraphics[scale=0.62]{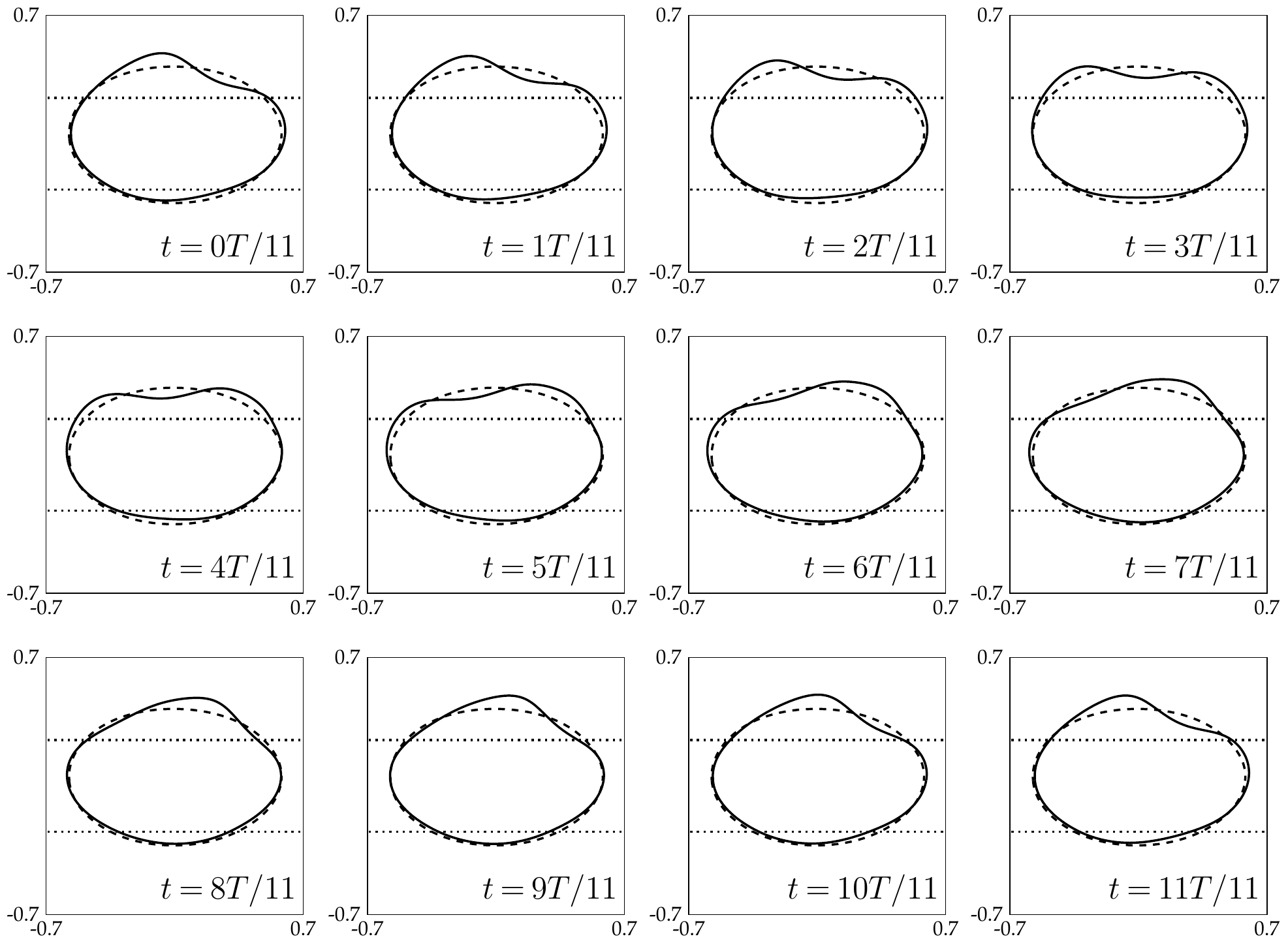}  
  \caption{\small{Stable periodic orbit near H2 over one period. The
      dashed shape is the fully nonlinear steady state
      solution. Bubble shapes for $Q^{-1} =
      Q_{2}^{-1}(1-\varepsilon^2)$, where $Q_2$ is the critical value
      for H2 and $\varepsilon=0.2$. The period is given by $T =
      2\pi/\omega$ where $\omega = \omega_c + \varepsilon^2(c - ad/b)$
      which in this case is $T = 0.5429$.}}
  \label{fig:periodic_orbits_h2_2}
\end{figure}

\begin{figure}
  \centering \includegraphics[scale=0.94]{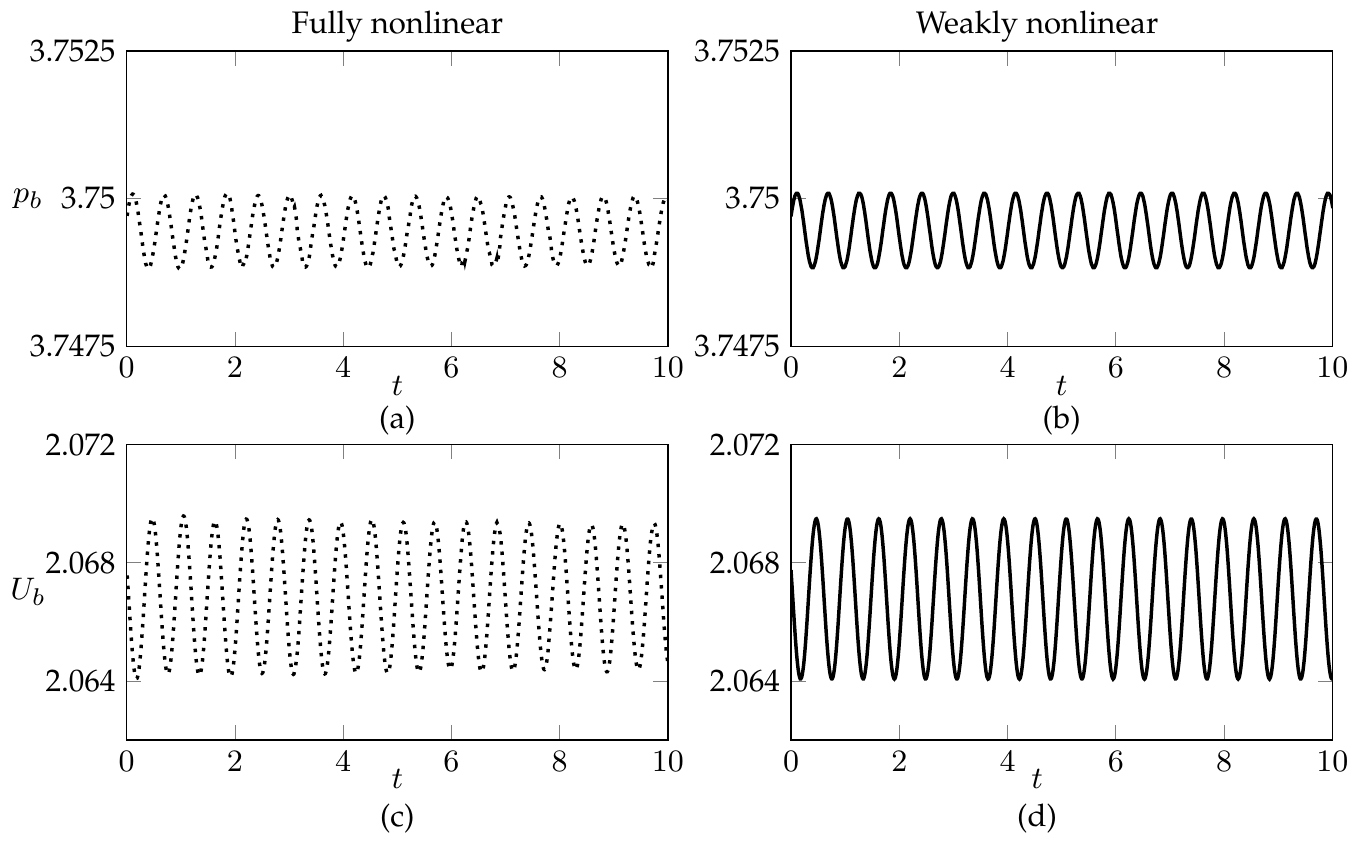}  
  \caption{\small{Comparison between the fully nonlinear
      time-dependent solutions and the weakly nonlinear solutions near
      H2. $Q^{-1} = Q_{2}^{-1}(1-\varepsilon^2)$ with
      $\varepsilon=0.01$. The dotted curves and solid curves are the
      fully nonlinear and weakly nonlinear solutions respectively. (a)
      and (b) The time signal of $p_b$. (c) and (d) The time signal of
      $U_b$.}}
  \label{fig:periodic_orbits_h2_1}
\end{figure}

An example of the weakly nonlinear bubble shapes of the periodic
orbits near the H2 point is shown in
figure~\ref{fig:periodic_orbits_h2_2}; now the underlying steady state
is asymmetric. The stable periodic orbit near the H2 point is
characterised by oscillations appearing on the single side of the
bubble that is over the edge of the depth-perturbation. These periodic
orbits are linearly stable and hence time-integration of the fully
nonlinear system should give a good indication of the period and
amplitude of the orbit as $t$ progresses. As the centroid of the
bubble shape does not vary significantly, more appropriate
time-signals are $p_b$ and $U_b$, shown in
figure~\ref{fig:periodic_orbits_h2_1}. The period and amplitude of the
weakly nonlinear periodic orbits demonstrate excellent agreement with
the fully nonlinear time-dependent simulations. 

\section{Conclusion}\label{sec:six}

We have investigated the transient dynamics and invariant solutions of
a finite air bubble propagating in a perturbed Hele-Shaw channel, a
model system in which to explore complex transition scenarios in fluid
mechanics. In
\S~\ref{sec:one}, time-dependent simulations confirmed that a number
of different modes of propagation are possible, including multi-tipped
solutions and oscillatory solutions as the flow-rate increases. These
simulations and an analysis of the steady solution space revealed a
finite width region of bistability (with respect to steady states). The demarcation of this bistable
region by two Hopf bifurcations explains the oscillatory solutions
observed from the time simulations. The first Hopf bifurcation, H1, is
subcritical, resulting in the emergence of unstable periodic orbits
and the second, H2, is supercritical, resulting in stable periodic
orbits. \JackRed{An interesting feature is that the nature of the
  bistability changes as $Q$ increases.  Initially ($Q$ immediately
  larger than $Q_1$) a slower symmetric steady state, S3, co-exists
  with an ever-present asymmetric steady state, AS1. A transition
  occurs at the pitchfork bifurcation as the stable AS2 asymmetric
  branch emerges. Finally due to the supercritical Hopf bifurcation,
  H2, a stable periodic orbit appears as an alternative stable
  invariant solution to the asymmetric AS1 branch.}

The criticality of the Hopf bifurcations was determined and periodic orbits were approximated using a weakly nonlinear
stability analysis in \S~\ref{sec:four}. 
\JackRed{The method described here applies to a range of governing
  equations and is independent of the choice of discretisation. It has
  been implemented in a general form in the open-source library
  \texttt{oomph-lib}}.
The approximate periodic orbits were in good agreement with the fully nonlinear
transient simulations. The oscillations of the unstable orbit near H1
manifest themselves as oscillations all the way around the bubble that
cause an oscillatory change in the centroid, $\overline{y}$, of the
bubble. For flow rates just above the H2 bifurcation point, stable
periodic orbits exist. The oscillations now appear as `waves'
on the side of the bubble above the depth-perturbation only, with the
lower edge of the bubble `resting' on the lower limits of the
depth-perturbation. In this case the offset of the bubble centroid
remains non-zero and constant and the underlying steady state is
asymmetric. These stable, asymmetric oscillations have similar
characteristics to the oscillations found for air-fingers by
\cite{de2009tube,pailha2012oscillatory,hazel2013multiple,thompson2014multiple}
and can be viewed as the finite-air bubble analogue. The
oscillations are driven by interactions between the edge of the bubble
and the edge of the constriction, which is typically only possible for
asymmetric bubble shapes.  In contrast the unstable periodic orbits
near H1 have not been seen before and appear to be a new phenomena in
this system. In this case the oscillations may be driven by
underdamping in the mechanism that restores a symmetric bubble to the
centre of the channel after perturbation; the generic mechanism 
requires an interaction between bubble deformation and surface tension
and is described in \cite{franco2017propagation}.

The periodic orbits near the H1 bifurcation point, although unstable,
are important in understanding the full time-dependent behaviour of
the system. Figure \ref{fig:periodic_orbits_h1_3}(a) is intriguing
evidence that these unstable periodic orbits are edge states of the
system (see, for example
\cite{kerswell2005recent,duguet2008transition}). Slight changes in the
initial condition can result in either the bubble returning to the
steady S2 state or breaking up/evolving to the AS1 solution. The
evidence presented here shows that the unstable periodic orbit is on
the boundary between these two forms of transient behaviour. We have
quantified the edge state by approximating the unstable periodic
orbits by a weakly nonlinear approximation but others have calculated
these by using edge-tracking techniques, including shear flow
\cite{kerswell2005recent,duguet2008transition}, and droplet pinch-off,
\cite{gallino2018edge} which rely on two distinct types of transient
behaviour being present. An added complication in this case is that
are three possible eventual outcomes; the bubble returning to the S2
steady state, breaking up, or evolving to the AS1 symmetric state. We
would expect to see evidence of the  periodic orbits and the
transitions in the bistable region in any experimental results.  The
periodic orbit near H1, despite being unstable, is therefore highly
influential in the underlying dynamics of a bubble and understanding
the basin boundary and a full comparison of the sensitivity of bubble
break up to initial conditions, especially close to bifurcation
points, is the subject of a combined experimental and theoretical
study currently underway. 

\section{Acknowledgments}

We acknowledge funding from the Engineering and Physical
Sciences Research Council through grant number EP/P026044/1.

\vskip2pc

\bibliographystyle{RS} 
\bibliography{hele_shaw_paper}

\end{document}